\documentclass[letterpaper,twocolumn,10pt]{article}
\usepackage{usenix-2020-09}

\usepackage{tikz}
\usepackage{amsmath}
\usepackage{caption}

\usepackage{graphicx}
\usepackage{soul}
\usepackage[many]{tcolorbox}
\usepackage{etoolbox}

\usepackage{dsfont}
\usepackage{mathtools,amsmath,amssymb,latexsym,amsfonts,stmaryrd}
\usepackage{pbox}
\usepackage{amsthm}
\usepackage{multirow}
\usepackage{xfrac}
\usepackage{tikz}
\usepackage{tkz-euclide}
\usepackage{hyperref}
\usepackage{mathrsfs}
\usepackage{bm}
\usepackage{mathpartir}
\usepackage{url}
\usepackage{syntax}
\usepackage{framed}
\usepackage[ruled,linesnumbered,vlined]{algorithm2e}
\usepackage[noend]{algpseudocode}
\usepackage{flushend}
\usepackage{listings}
\usepackage{moresize}
\usepackage{wrapfig}
 
\usepackage{seqsplit}
\usepackage{alltt}
\usepackage{xspace}
\usepackage{enumitem}
\usepackage{xcolor}
\usepackage{pifont}

\DeclareMathAlphabet{\mathcal}{OMS}{cmsy}{m}{n}

\usepackage{amsmath, listings, amsthm, amssymb, proof, xspace}

\usepackage{thmtools}

\declaretheoremstyle[spaceabove=\topsep,notefont=\normalfont\itshape]{mystyle}

\newcommand{\revise}[2]{{\color{red}{\ifx&#1&\else- #1\fi}} {\color{ForestGreen}{\ifx&#2&\else+ #2\fi}}}%
\renewcommand{\revise}[2]{#2}%

\usetikzlibrary{arrows,patterns, decorations.pathreplacing}

\newcommand{\Appx}{Appx.}
\newcommand{\F}{Fig.}
\newcommand{\E}{Eq.}
\newcommand{\T}{Table}
\renewcommand{\S}{Sec.}
\newcommand{\A}{Alg.}

\newcommand{\parh}[1]{\noindent\textbf{#1}}
\newcommand{\parhs}[1]{\noindent\underline{\textit{#1}}}

\newcommand{\Lem}{Lemma}
\newcommand{\Thm}{Theorem}

\newcommand{\Df}{Def.}

\newtheorem{theorem}{Theorem}

\newtheorem{lemma}{Lemma}
\newtheorem{definition}{Definition}

\newcommand{\ignore}[1]{}

\lstdefinestyle{base}{
  moredelim=**[is][\color{red}]{@}{@},
  escapeinside={<@}{@>}
}

\newcommand{\tool}{\textsc{GCert}\xspace}

\newcommand{\ms}{$M_s$\xspace}
\newcommand{\mc}{$M_c$\xspace}
\newcommand{\ma}{$M_a$\xspace}

\usepackage{amssymb}
\usepackage{pifont}
\usepackage{bbding}

\newcommand\DejaVuttfamily{%
  \fontfamily{DejaVuSansMono-TLF}\selectfont }

\lstdefinestyle{base}{
  moredelim=**[is][\color{red}]{@}{@},
  escapeinside={<@}{@>}
}

\lstdefinelanguage
   [x64]{Assembler}     
   [x86masm]{Assembler} 
   {morekeywords={CDQE,CQO,CMPSQ,CMPXCHG16B,JRCXZ,LODSQ,MOVSXD, %
                  POPFQ,PUSHFQ,SCASQ,STOSQ,IRETQ,RDTSCP,SWAPGS, %
                  rax,rdx,rcx,rbx,rsi,rdi,rsp,rbp, %
                  r8,r8d,r8w,r8b,r9,r9d,r9w,r9b}} 

\let\oldbibliography\thebibliography
\renewcommand{\thebibliography}[1]{%
  \oldbibliography{#1}%
  \setlength{\itemsep}{0pt}%
}

\usepackage{listings}
\usepackage{fancyvrb}
\usepackage{color}
\definecolor{lightgray}{rgb}{.9,.9,.9}
\definecolor{darkgray}{rgb}{.4,.4,.4}
\definecolor{purple}{rgb}{0.65, 0.12, 0.82}
\definecolor{commentgreen}{RGB}{63,127,95}
\definecolor{pptred}{RGB}{192,0,0}
\definecolor{pptgreen}{RGB}{226,240,217}
\definecolor{pptyellow}{RGB}{255,242,204}
\definecolor{pptblue}{RGB}{222,235,247}
\definecolor{pptorg}{RGB}{244,177,131}
\definecolor{pptgreen1}{RGB}{31,78,121}
\definecolor{pptgreen2}{RGB}{0,153,153}

\newlength{\dpcircle}
\newlength{\rcircle}
\newlength{\dcircle}
\newcommand{\docircle}[4]{%
  \setlength{\dpcircle}{\dp\strutbox}%
  \setlength{\dcircle}{\dpcircle}%
  \addtolength{\dcircle}{\ht\strutbox}%
  \setlength{\rcircle}{0.5\dcircle}%
  \setlength{\unitlength}{1sp}%
  \begin{picture}(\number\dcircle,0)
    \color{#1}
    \put(\number\rcircle,\number\dpcircle){\circle*{\number\dcircle}}
    \color{#2}
    \put(\number\rcircle,\number\dpcircle){\circle{\number\dcircle}}
    \put(\number\rcircle,0){\makebox[0pt]{\textcolor{#3}{#4}}}
  \end{picture}%
}

\usepackage{threeparttable}
\colorlet{myPurple}{blue!40!red}
\definecolor{myOrange}{RGB}{255,192,0}

 \lstdefinelanguage{Solidity}{
   keywords={len,delete,int,void,payable, public, event, contract, typeof, new, true, false, catch, function, return, null, catch, switch, var, if, in, while, do, else, case, break,unsigned,int32_t,int16_t,for,define},
   keywordstyle=\color{violet}\bfseries,
   ndkeywords={State,Mem,i64,i32,i8, PC},
   ndkeywordstyle=\color{ForestGreen}\bfseries,
   identifierstyle=\color{black},
   sensitive=false,
   comment=[l]{//},
   morecomment=[s]{/*}{*/},
   commentstyle=\color{commentgreen}\ttfamily,
   stringstyle=\color{red}\ttfamily,
   morestring=[b]',
   morestring=[b]"
 }

\newcommand{\rnum}[1]{\uppercase\expandafter{\romannumeral #1\relax}}

\algnewcommand{\LeftComment}[1]{\Statex \(\triangleright\) #1}

\NewDocumentCommand{\statcirc}{ O{#2} m }{%
    \begin{tikzpicture}
    \fill[#2] (0,0) circle (0.8ex); 
    \fill[#1] (0,0) -- (180:0.8ex) arc (180:0:0.8ex) -- cycle; 
    \end{tikzpicture}
}

\tikzset{%
    pics/sema/.style args={#1/#2/#3}{code={%
        \ifstrequal{#2}{0}{%
            \node[circle,minimum width=1.4mm,draw,fill=#1] {};
        }{%
            \tkzDefPoint(0,0){O}
            \tkzDrawSector[R,fill=#1](O,1.2mm)(90,90-#2)
            \tkzDrawSector[R,fill=#3](O,1.2mm)(90-#2,90-360)
    }
    }},
}

\newcommand{\cBrush}{\textcolor{green}{\ding{51}}}
\newcommand{\xBrush}{\textcolor{red}{\ding{55}}}

\newcommand{\Line}[2]{\hyperref[#2]{\textcolor{darkgray}{L\texttt{#1}}}}
\newcommand{\cirC}[2][white]{{\footnotesize\docircle{#1}{black}{black}{#2}}}

\newcommand{\mydiamond}[1]{%
  \sbox0{$\lozenge$}%
  \usebox0\kern-.5\wd0\clap{\raisebox{.1ex}{\scalebox{.7}[1]{#1}}}\kern.5\wd0%
}

\newcommand{\norm}[1]{\left\lVert#1\right\rVert}
\newcommand{\dif}[1]{\mathrm{d}#1}

\DeclareMathOperator*{\argmin}{arg\,min}

\usepackage{etoolbox}
\makeatletter
\patchcmd{\maketitle}
	{\@maketitle}
	{\@maketitle\vspace{-10pt}}
	{}
	{}
\makeatother

\let\OLDthebibliography\thebibliography
\renewcommand\thebibliography[1]{
  \OLDthebibliography{#1}
  \setlength{\parskip}{0pt}
  \setlength{\itemsep}{0pt plus 0.3ex}
}

\begin{document}

\title{Precise and Generalized Robustness Certification for Neural Networks
\thanks{The extended version of the USENIX Security 2023 paper~\cite{yuan2023precise}.}}

\author{
{\rm $^{1,2}$Yuanyuan Yuan\thanks{This work is done when Yuanyuan Yuan was visiting ETH Zurich.},\; $^{1}$Shuai Wang\thanks{Corresponding author.},\; and $^{2}$Zhendong Su}\\
$^{1}$The Hong Kong University of Science and Technology, \; $^{2}$ETH Zurich\\
\textit{yyuanaq@cse.ust.hk, \; shuaiw@cse.ust.hk, \; zhendong.su@inf.ethz.ch}
}

\maketitle

\begin{abstract}
  The objective of neural network (NN) robustness certification is to
  determine if a NN changes its predictions when mutations are made to its
  inputs. While most certification research studies pixel-level or a few
  geometrical-level and blurring operations over images, this paper proposes a
  novel framework, \tool, which certifies NN robustness under a precise and
  unified form of diverse semantic-level image mutations. 
  We formulate a comprehensive set of semantic-level image mutations
  \textit{uniformly} as certain directions in the latent space of generative
  models. We identify two key properties, \textit{independence} and
  \textit{continuity}, that convert the latent space into a precise and
  analysis-friendly input space representation for certification. \tool\ can be
  smoothly integrated with de facto complete, incomplete, or quantitative
  certification frameworks. With its precise input space representation, \tool\
  enables for the first time complete NN robustness certification with moderate
  cost under diverse semantic-level input mutations, such as weather-filter,
  style transfer, and perceptual changes (e.g., opening/closing eyes). We show
  that \tool\ enables certifying NN robustness under various common and
  security-sensitive scenarios like autonomous driving.
\end{abstract}

\section{Introduction}
\label{sec:introduction}

While neural networks (NNs) have prosperous development, their robustness issue
remains a serious concern. Specifically, a NN's prediction can be easily changed
by applying small mutations over its input (e.g., via adversarial
attacks~\cite{goodfellow2015explaining,hendrycks2021natural,pei2017deepxplore}).
This has led to severe outcomes in security-critical applications like 
autonomous driving~\cite{tesla,report}.

As a principled solution to mitigate the robustness issue and adversarial attacks,
NN robustness certification has been actively
studied~\cite{gehr2018ai2,bonaert2021fast,balunovic2019certifying,singh2019abstract,
lecuyer2019certified,lorenz2021robustness,alfarra2022deformrs}. Given a mutation $\tau$ and a tolerance $\| \delta_{\max}
\|$ (which decides the maximum extent to which the mutation is applied) over an
input image, it aims to certify that the NN's prediction for all mutated images
remains within the tolerance. The primary challenge is to \textit{formulate the
input space} determined by $\tau$ and $\| \delta_{\max} \|$ in a precise and
analysis-friendly manner. It is evident that the input space has an
\textit{infinite} number of mutated images, and in practice, various input space
abstraction schemes are applied, e.g., using the interval and zonotope abstract
domains originated from the abstract interpretation
theory~\cite{cousot1977abstract}. Technically, verifying NNs can be both sound
and complete (NNs are generally loop-free)~\cite{xu2020fast,wang2018formal,
ferrari2021complete,wang2021beta,jordan2019provable}. 

In practice, de facto NN certification delivers \textit{sound} analysis, while
most of them fail to provide \textit{complete} analysis simultaneously. This is
primarily due to the over-approximated (imprecise) input space formed over
diverse real-world mutations.
Most certification works focus on pixel-level input mutations (e.g., adding
$\ell_p$-norm noise) whose resulting input space is linear and easy for analysis.
Nevertheless, pixel-level mutations are simple and insufficient to subsume
various mutations that may occur in reality. Note that recent testing and
adversarial attacks toward NNs have been using geometrical
mutations~\cite{xie2019deephunter,xiao2018spatially} (e.g., rotation),
blurring~\cite{xie2019deephunter,tian2018deeptest}, and more advanced mutations including weather
filter~\cite{zhang2018deeproad,yuan2022unveiling}, style transfer~\cite{geirhos2018imagenet,yuan2022revisiting,yuan2022unveiling}, and
perceptual-level (e.g., changing hairstyle)
mutations~\cite{dunn2021exposing,yuan2021enhancing,yuan2021perception}.

It is challenging to take into account the aforementioned mutations, because the
input spaces formed by applying those mutations over an image are
\textit{non-linear}, making a precise input space representation difficult.
Moreover, some mutations even do not have explicit mathematical expressions.
Recent research considers geometrical mutations, but over-approximates the input
space~\cite{singh2019abstract,balunovic2019certifying,mohapatra2020towards}. As
a result, it can only support incomplete certification, which may frequently
fail due to the over-approximation (``false positives'') rather than
non-robustness (``true positives''). Other works support blurring, by only
offering (incomplete) probabilistic certification: a NN is certified as
``robust'' with a
probability~\cite{li2021tss,hao2022gsmooth,fischer2020certified}. A
probabilistic guarantee is less desirable in security-critical scenarios. Also,
existing techniques are typically \textit{mutation-specific}; supporting new
mutations is often challenging and costly, and may require domain expertise.

Besides pixel-level mutations, adversarial NN attacks also adopt
\textit{semantic-level mutations}, which change image semantics holistically.
Generative models (e.g.,
GAN~\cite{goodfellow2020generative}, VAE~\cite{kingma2014auto}), which map
points from a latent space to images, can generate infinite images with diverse
semantics. By using the latent space of generative models and their image
generation capability, we explore a principled technical solution of forming
input space representations of various semantics-level mutations to support NN
certification.

Conceptually, mutating an image $x$ via a generative model is achieved by moving
its corresponding latent point $z$ in the latent space. However, given the
latent spaces of common generative models, mutating $x$ would only generate
mutated images $x'$ with arbitrarily changed content. Thus, we re-formulate the
latent spaces of generative models in a \textit{regulated} form (see below),
thereby offering a \textit{precise} and \textit{analysis-friendly} input space
representation to support complete NN certification. We re-form
the latent spaces with the aim of achieving two key objectives: 1)
\textit{independence}: image semantics corresponding to different mutations
change independently, and 2) \textit{continuity}: if certain semantics are
mutated, they should change continuously. With independence, different
semantic-level mutations are represented as orthogonal directions in the latent
space, and mutating a semantics instance (e.g., hair style) is conducted by
moving a point along the associated direction in the latent space without
disturbing other semantics. Moreover, continuity ensures latent points
of all mutated inputs are exclusively included in a segment in the latent space.

Our analysis for the first time incorporates a broad set of semantic-level
mutations, including weather filter (e.g., rainy), style transfer, and
perceptual mutations, to deliver comprehensive and practical NN certification.
Our approach is agnostic to specific generative models and enables
certification towards mutations that have no explicit math expression. It offers
precise input space representations, and consequently
enables the first complete certification for semantic-level mutations with
largely reduced cost (from exponential to polynomial). Also, since mutated
inputs are represented as a segment in the latent space (see
\S~\ref{sec:overview}), our approach can be integrated into recent quantitative
certification frameworks~\cite{mirman2021robustness}, offering fine-grained
bounds to depict NN robustness.

We implement our approach in a framework named \tool. We conduct both
qualitative and quantitative evaluations to assess the diversity and correctness
of semantic-level mutations enabled in \tool. Our precise input space
representation is around $60\%$ tighter than SOTA results for geometrical
mutations. Moreover, we show that \tool\ can be smoothly leveraged to boost de
facto complete, incomplete, or quantitative certification frameworks toward
various NNs  
under usage scenarios like classification, face recognition, and autonomous
driving. We also cross-compare how different mutations can influence NN
robustness. We then present both theoretical and empirical cost analysis for
\tool, illustrating its applicability. Overall, we have the following
contributions: 

\begin{itemize}[noitemsep,topsep=0pt]
  \item This work enables the first sound and complete NN certification under
  various real-world semantic-level mutations, which is a major step toward
  practical NN robustness certification. 
  We uniformly re-formulate a broad set of semantic-level image mutations as
  mutations along directions in the latent space of generative models. Moreover,
  this work for the first time certifies NN robustness under advanced mutations,
  such as weather-filter, style transfer, and perceptual mutations.

  \item We identify two key requirements, independence and continuity, that
  turn the latent space associated with common generative models into a
  \textit{precise} and \textit{analysis-friendly} form for NN certifications. We give
  formal analysis and practical solutions to achieve these requirements.
  Consequently, input spaces under semantic-level mutations can be represented
  precisely, for the first time enabling practical complete/quantitative certification.

  \item Evaluations illustrate the correctness and diversity of our delivered
  semantic-level mutations. Our approach manifests high practicality and
  extendability, as it can be applied on any common generative models and
  smoothly incorporated into existing certification frameworks.
  We certify different NNs under various security-sensitive scenarios and
  comprehensively study NN robustness w.r.t. the NN structure, training data,
  and input mutations.
\end{itemize}

\parh{Research Artifact.}~We release
the code and data at
\url{https://github.com/Yuanyuan-Yuan/GCert}~\cite{snapshot}.

\section{Preliminary \& Background}
\label{sec:background}

In this section, we first introduce NN robustness certification, and then
discuss how various input mutations are performed. We then briefly
review existing input representation solutions and their limitations to
motivate this research. 

\subsection{NN Robustness Certification}
\label{subsec:problem}

Following existing works~\cite{gehr2018ai2,zhang2018efficient,bonaert2021fast,
li2021tss,muller2022prima,lecuyer2019certified,ferrari2021complete}, we focus on
\textit{robustness} certification of image classification, a common and essential
NN task. Certifying other NN tasks, objectives (e.g., fairness),
and types of inputs (e.g., audio, text) can be easily extended from \tool;
see \S~\ref{sec:discussion}.

\begin{figure}[!ht]
    \captionsetup{font=footnotesize}
    \centering
    \vspace{-5pt}
    \includegraphics[width=0.98\linewidth]{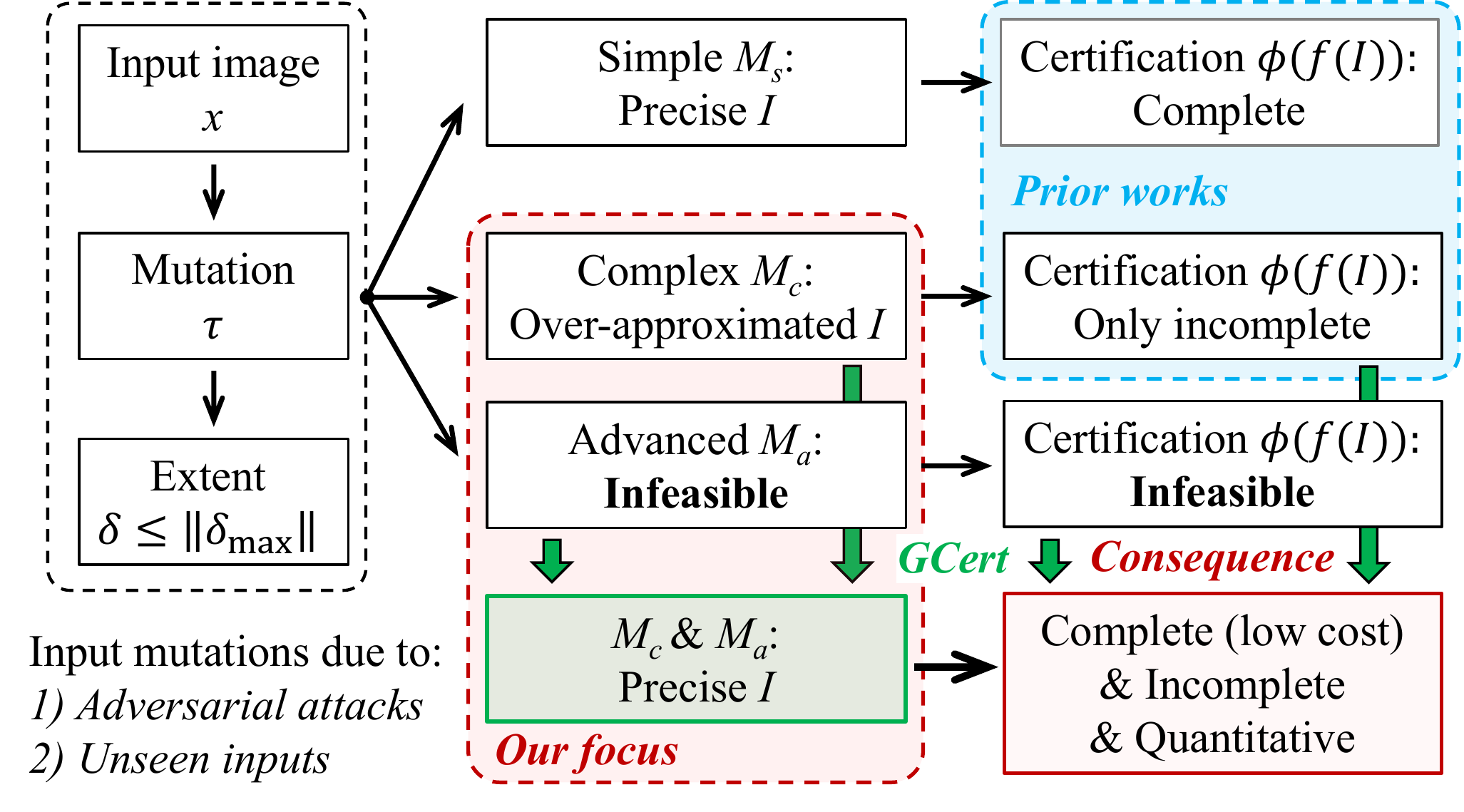}
    \vspace{-10pt}
    \caption{Certification procedures and the position of \tool.}
    \vspace{-10pt}
    \label{fig:position}
\end{figure}

\parh{Problem Statement.}~As illustrated in \F~\ref{fig:position}, NN robustness
certification comprises the following components: 1) a certification framework
$\phi$, 2) a target NN $f$, 3) an input image $x$, 4) a mutation scheme
$\langle \tau, \delta \rangle$, where $\tau$ is a function specifying the
mutation over $x$ and $\delta$ decides to what extent the mutation is performed
(e.g., $\delta$ denotes the rotating angle if $\tau$ is rotation), and 5)
tolerance $\| \delta_{\max} \|$, specifying the maximal value (often in the form of norm
since $\delta$ can be a vector) of $\delta$. 

\parh{Threat Model.}~Our threat model is aligned with existing works in this
field~\cite{gehr2018ai2,zhang2018efficient,
bonaert2021fast,li2021tss,muller2022prima,lecuyer2019certified,ferrari2021complete}.
That is, we aim to certify that when one mutation scheme $\langle \tau, \delta
\rangle$, which is bounded by $\| \delta_{\max} \|$ (i.e., $0 \leq \| \delta \|
\leq \| \delta_{\max} \|$), is applied on an image $x$, whether the NN $f$ has
the same prediction with $x$ over \textit{all} mutated images.
As will be detailed in \F~\ref{fig:position}, the mutations can be introduced by
variations in real-world unseen inputs or attackers who are generating
adversarial examples (AEs)~\cite{goodfellow2018making} using different methods.
A certified NN is robust to those unseen or adversarial inputs, and we do not
assume any particular AE generation method.
Formally, we study if the following condition holds.
\begin{equation}
\label{equ:certification}
    f(x) = f(x'),
    \;
    \forall x' \in \{ \tau(x, \delta) \; |  \; 0 \leq \| \delta \|
                                          \leq \| \delta_{\max} \| \}
\end{equation}

\noindent For instance, if $\tau$ is rotation and $\| \delta_{\max} \|$ is set
to $30^{\circ}$, the certification aims to certify whether $f$ changes its
prediction for $x$ if $x$ is rotated within $30^{\circ}$. A certification is
successful if it certifies that \E~\ref{equ:certification} always holds; it
fails otherwise. We now introduce input mutation and the certification procedure.

\parh{Abstracting Mutated Inputs.}~Typical challenges include 1) considering
representative real-world image mutations, and 2) forming an input space
representation $I$ which should be as close to all $x' = \tau(x, \delta)$ as
possible~\cite{balunovic2019certifying,li2021tss,fischer2020certified,
hao2022gsmooth,pautov2022cc}. Image mutations $\tau$ are typically implemented
in distinct and often challenging ways (see \S~\ref{subsec:transformation}), and
therefore, previous works are usually specific to one or a few mutations that
are easier to implement and less costly. Even worse, the input space formed by
$\tau(x, \delta)$ often captures unbounded sets of images, and one cannot simply
enumerate all images $\tau(x, \delta)$ and check \E~\ref{equ:certification}.
In practice, various abstractions are performed over $\tau(x, \delta)$ to offer
an analysis-friendly representation $I$ (which are generally imprecise; see
below). For instance, the interval $[a, b]$ can be viewed as an abstraction of a
pixel value $v$ mutated within a bound $a \leq v \leq b$. Nevertheless, $\tau(x,
\delta)$ resulted from most mutation schemes are non-linear, making a proper
$I$ (without losing much precision) fundamentally challenging.

\parh{Certification Procedure.}~Given input space representation $I$,
certification propagates it through NN layers to obtain outputs, which
are then checked against \E~\ref{equ:certification}.
Due to Rice's theorem, constructing a sound and complete certification
(verification) over non-trivial program properties is
undecidable~\cite{rice1953classes}, and most verification tools are
unsound~\cite{livshits2015defense}. Nevertheless, for NNs (which are generally
loop-free), it is feasible to deliver both sound and complete certification
(e.g., with solver-based approaches)~\cite{wang2018formal,
wang2021beta,ferrari2021complete}; the main challenge is scalability with the
size of NNs (e.g., \#layers). In fact, de facto certification tools consistently
offer sound analysis, such that when the NN $f$ violates a property under an input
$x' \in I$, the certification can always reveal that (i.e., no false negative).

On the other hand, the ``completeness'' may not be guaranteed: the certification is
deemed complete if it proves \E~\ref{equ:certification} holds when it actually
holds (i.e., no false positive). Nevertheless, complete
certifications~\cite{ferrari2021complete,wang2021beta,zhang2018efficient,
wang2018formal,xu2020fast} are often challenging to conduct. A
complete certification procedure requires the input space representation $I$
must be \textit{precise}; i.e., $I$ should \textit{exclusively} include \textit{all}
mutated inputs $x' \in \tau(x, \delta)$. To date, complete certifications are
limited to a few pixel-level mutations~\cite{wang2018formal,ferrari2021complete,wang2021beta,li2023sok}
and the overall cost is exponential to the neuron width (the maximum \#neurons in a layer).

Incomplete certifications~\cite{muller2022prima,singh2019abstract,gehr2018ai2}
gradually over-approximate the input representation along the propagation to
reduce the modeling complexity. An incomplete certification, however, cannot
guarantee that it can prove \E~\ref{equ:certification} which actually holds.
Hence, we may often encounter false positives, treating a safe NN as
``unknown/unsafe.''

\vspace{-5pt}
\subsubsection*{Design Focus \& Position w.r.t. Previous Works}
\vspace{-5pt}

A diverse set of real-world image mutation schemes (see \S~\ref{subsec:transformation})
are used in attacking NNs. These mutations are
holistically categorized as simple \ms, complex \mc, and advanced \ma.
As shown in \F~\ref{fig:position}, most prior works employ \ms\ (e.g.,
adding noise bounded by $\ell_p$-norm) whose input
space is linear and putting primary efforts into optimizing the certification
frameworks. Existing complete certifications are limited to simple mutations.
\mc's input space is non-linear and over-approximated, and therefore,
certifications based on \mc\ are only incomplete. \ma\ was infeasible, since
they do not have ``explicit'' math expressions. It is thus unclear how to
represent their resulting input space.

The design focus of \tool\ is orthogonal to previous works: it addresses a
fundamental and demanding challenge, i.e., how to offer a \textit{precise} and
\textit{unified} input space representation $I$ over a wide range of real-world
semantic-level mutations $\tau$ (i.e., those belonging to \mc\ and \ma). The
abstracted $I$ can be integrated with de facto certification frameworks, and
consequently enables sound, complete, and even quantitative (which quantifies
the percentage of mutated inputs that retain the prediction) NN certification
with moderate cost.

\subsection{Real-World Diverse Image Mutations}
\label{subsec:transformation}

\begin{figure}[!ht]
    \captionsetup{font=footnotesize}
    \centering
    \vspace{-5pt}
    \includegraphics[width=0.90\linewidth]{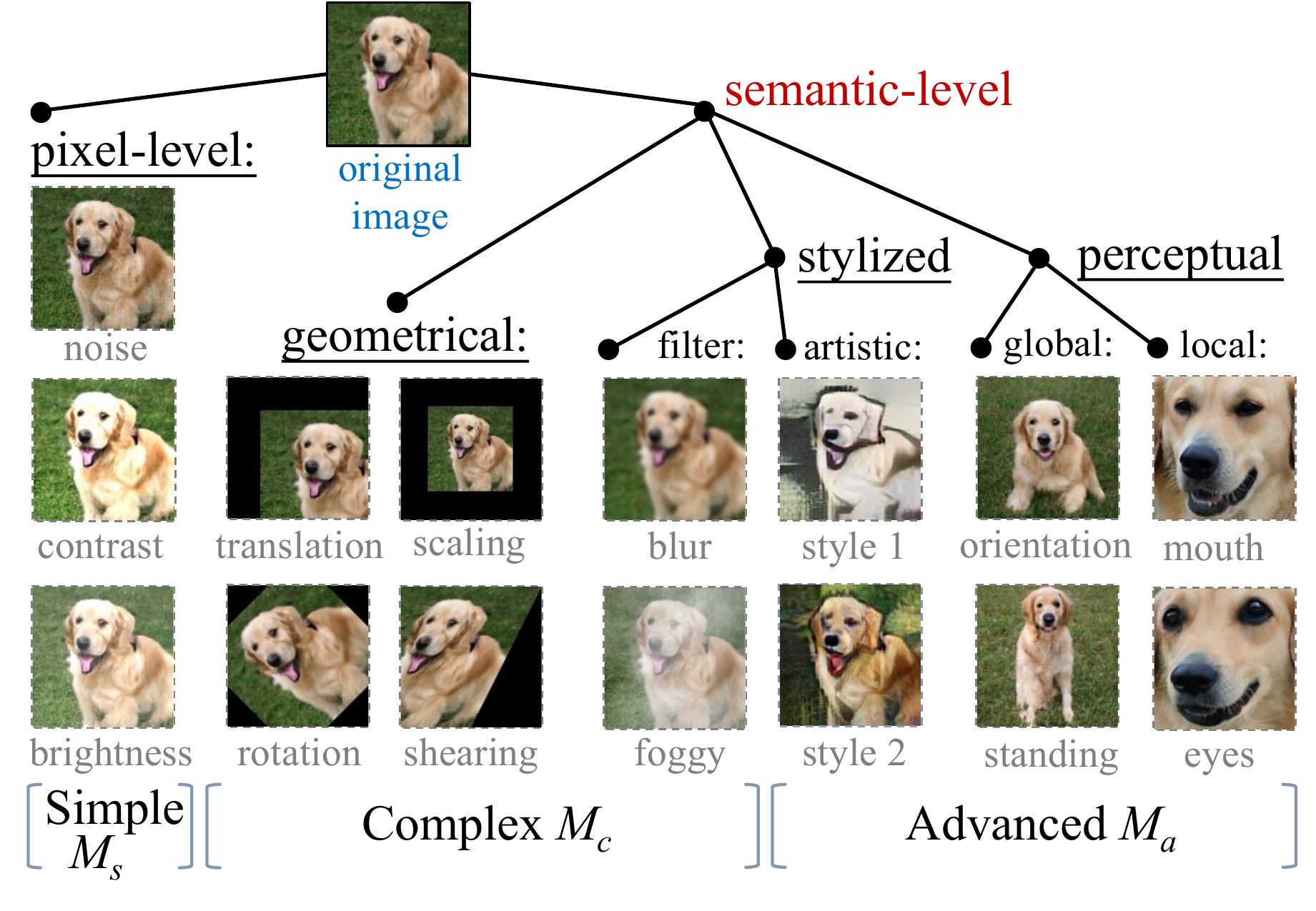}
    \vspace{-10pt}
    \caption{Hierarchical categorization of image mutations. Geometrical,
    stylized, and perceptual mutations belong to semantic-level because they
    holistically change semantics of the original image. \tool enables all
    semantic-level mutations (\mc\ and \ma) and offers an analysis-friendly,
    linear, and unified representation for complete certification.}
    \vspace{-10pt}
    \label{fig:mutation}
\end{figure}

\F~\ref{fig:mutation} hierarchically categorizes image mutations that have been
adopted in testing and attacking NNs by prior works. We introduce each category
below.

\parh{Simple \ms.}~Mutations of this category directly operate on image pixels.
Given an input image $x$, one \ms\ can be expressed as either $x + \delta$, or
$x \times \delta$. For example, to change image brightness (or contrast), a real
number is added to (or multiplied with) all pixel values. Similarly, to add noise,
a vector $\delta$ of random values is added on $x$. 
Overall, pixel values of mutated images change \textit{linearly} with the norm
of $\delta$. Therefore, an input space $I$ corresponding to these
mutations can be precisely formed and propagated along NN layers.

\begin{figure}[!ht]
    \captionsetup{font=footnotesize}
    \centering
    \vspace{-10pt}
    \includegraphics[width=0.9\linewidth]{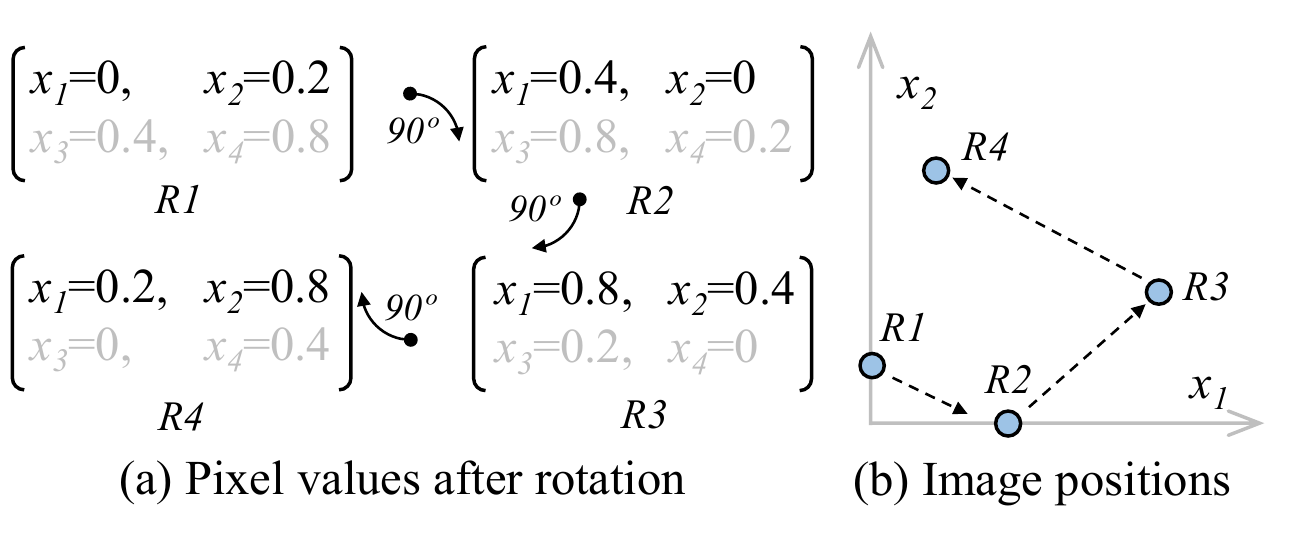}
    \vspace{-10pt}
    \caption{Positions of rotated images. In (a), a four-pixel image $R1$ is
    right rotated $90^{\circ}$ three times into $R2$, $R3$, and $R4$. Their
    positions, when only considering the first two pixels, are shown in (b).
    Dashed arrows only point the direction where rotating angle is increased;
    intermediate images between two marked images are omitted in (b). }
    \label{fig:rotation}
    \vspace{-10pt}
\end{figure}

\parh{Complex \mc\ --- Geometrical.}~These mutations (a.k.a. affine
mutations) change geometrical properties (e.g., position, size) but preserve
lines and parallelism (i.e., move points to points) of input images. For any
pixel in the $(i, j)$-th location (which is a relative location to the centering
point) of an image $x$, geometrical mutations first compute $[i', j']^\intercal
= \bm{A} \times [i, j]^\intercal + \bm{b}$, where $\bm{A}$ is a $2 \times 2$
matrix and $\bm{b}$ is a 2-dimensional vector. Different $\bm{A}$ and $\bm{b}$
specify different geometrical schemes. For instance, if $\tau$ is rotation, then
\begin{equation*}
    \scriptsize
\begin{bmatrix}
    i' \\
    j' \\
\end{bmatrix} =
\begin{bmatrix}
    \cos \delta & - \sin \delta \\
    \sin \delta & \cos \delta \\
\end{bmatrix} \times
\begin{bmatrix}
    i \\
    j \\
\end{bmatrix} +
\begin{bmatrix}
    0 \\
    0 \\
\end{bmatrix} =
\begin{bmatrix}
    i \cos \delta - j \sin \delta \\
    i \sin \delta + j \cos \delta \\
\end{bmatrix},
\end{equation*}

\noindent where $\delta$ is the rotating angle ($\delta > 0$ for left
rotation). Each pixel in the mutated image $x'$ is decided as $x'_{i',j'} =
x_{i, j}$. Since $i'$ and $j'$ should be integers, interpolation is further
applied on pixels around $i'$ and $j'$ if they are
non-integer~\cite{balunovic2019certifying,geometric,li2021tss}.
\F~\hyperref[fig:rotation]{\ref*{fig:rotation}(a)} reports the pixel
values\footnote{Pixel values are 8-bit integers in most image formats (e.g., JPEG).
However, modern NN frameworks usually convert pixel values into floating-point numbers
in $[0, 1]$. We follow the latter setup in the rest of this paper.} of a four-pixel
image when it is right rotated $90^{\circ}$ three times.
\F~\hyperref[fig:rotation]{\ref*{fig:rotation}(b)} visualizes the first two
pixels across three mutations. The correlation between pixel values and $\delta$
is apparently \textit{non-linear} in geometrical mutations, which impedes
presenting a precise input space $I$ for geometrical mutations.

\parh{Complex/Advanced \mc/\ma\ --- Stylized.}~As in \F~\ref{fig:mutation}, both
filter-based and artistic mutations change an image's ``visual style.''
Filter-based mutations first define a domain-specific filter (such as blurring,
rainy, or foggy), and then apply the filter to an image to simulate varying
real-world scenarios. For example, as often in auto-driving 
testing~\cite{tian2018deeptest,zhang2018deeproad}, a driving scene image is applied by a foggy filter to
stress the NN for making correct driving decision. Artistic mutations perform
more extensive mutations to largely change the color scheme of an image. Recent
studies show that artistic-stylized images reveal the texture-bias of NNs and
effectively flip their predictions~\cite{geirhos2018imagenet,hermann2020origins}. Nevertheless, artistic
stylizing is never explored in NN robustness certification.

\parh{Advanced \ma\ --- Perceptual.}~Perceptual mutations aim to change
perceptual properties of input images, e.g., making a lying dog stand up (as
shown in \F~\ref{fig:mutation}). Holistically, its effectiveness 
relies on the manifold
hypothesis~\cite{bengio2013representation,zhu2018image}, which states that
perceptually meaningful images lie in a low-dimensional manifold that encodes
their perceptions (e.g., facial expressions, postures). In practice, generative
models are widely used for approximating manifold, such that they can generate
infinite and diverse mutated images after being trained on a finite number of
images. Intuitively, for dog images, since real-world images subsume various
perceptions, generative models can infer a posture mutation, ``standing
$\rightarrow$ lying'', based on existing standing and lying dogs. The trained
generative model then applies the mutation on other (unseen) images. Although perceptual
mutations facilitate assessing NNs for more advanced properties, they are hard
to control (e.g., several perceptions may be mutated together). Moreover,
perceptual mutations result in non-smooth mutations, e.g., the mutated image is
largely altered with a small perturbation on $\delta$. These hurdles make them
less studied in NN certification.

\subsection{Input Space Representation $I$}
\label{subsec:space}

As noted in \S~\ref{subsec:problem} and \F~\ref{fig:mutation}, existing NN certifications only
support \ms\ and \mc\ mutations. Below, we review existing input representations
$I$ and discuss their limitations.

\parh{Precise $I$ for \ms.}~Early works primarily consider pixel-wise mutations,
due to their simpler math forms and the derived \textit{linear} input space.
\ms\ primarily focuses on adding noise, and $\| \delta \|$, accordingly, is
computed using the $\ell_{p}$-norm:
\begin{equation*}
    \scriptsize
\label{equ:lp}
    \| \delta \|_{p} = (\delta[0]^p + \delta[1]^p + \dots + \delta[m-1]^p)^{\frac{1}{p}},
    \; \delta \in \mathbb{R}^m.
\end{equation*}
As noted in \S~\ref{subsec:transformation}, input space $I$ derived
from \ms\ can be precisely represented, i.e., $I$ exclusively includes all
mutated inputs. Current complete certifications only support \ms.

\parh{Imprecise $I$ for \mc.}~An explicit and rigorous math form may be
required to form input space $I$ over each \mc\ scheme. Hence, prior
certification works only consider a few geometrical/filter-based mutations.
These works are classified as \textit{deterministic} and \textit{probabilistic},
depending on if the certification proves that \E~\ref{equ:certification} is
always or probabilistically holds.

\parhs{Deterministic.}~DeepPoly~\cite{singh2019abstract} depicts the input space
of geometrical transformations as intervals, which is imprecise and frequently
results in certification failure~\cite{balunovic2019certifying}. Though it can
be optimized by Polyhedra relaxation, Polyhedra relaxation may hardly scale to
geometric transformations~\cite{balunovic2019certifying}, because its cost grows
exponentially with the number of variables. Also, the represented input space
with Polyhedra relaxation is still imprecise.
DeepG~\cite{balunovic2019certifying} computes linear constraints for pixel
values under a particular geometrical mutation. It divides possible
$\delta$ values into splits before sampling from these small splits to obtain
unsound constraints (which miss some mutated images) for pixel values. Later,
based on the maximum change in pixel values, it turns unsound constraints into
sound ones by including all mutated images but also extra ones.
\tool, however, can precisely represent the input space $I$
with a polynomial cost.

GenProve~\cite{mirman2021robustness} uses generative models to enhance
certification. However, the generative model is used for generating
interpolations between the original image and the mutated image. Several
problems arise: 1) for advanced, perceptual mutations (not
considered in GenProve), the target image for interpolation may not always be
accessible, e.g., for a dog, we must have two images where it is standing in one
and lying in another; 2) interpolations are not guaranteed to only follow the
semantic differences between the original image and the mutated image, e.g.,
interpolated images between faces of different orientations may have gender
changed as well; and 3) interpolations are not always ``intermediate images''
between the original and mutated ones, e.g., when interpolating between an image
and its $30^{\circ}$-rotated mutant using generative model, a
$90^{\circ}$-rotated image may exist in the interpolations.

\parhs{Probabilistic.}~Other works apply randomized smoothing on the target NN
to support more mutations (e.g., filter-based mutations) and improve scalability~\cite{li2021tss,fischer2020certified,hao2022gsmooth}.
In general, different smoothing strategies are tailored to accommodate mutations of
different mathematical properties (e.g., translation,
rotation~\cite{li2021tss}). However, due to smoothing,
\E~\ref{equ:certification} can only be \textit{probabilistically} certified, if
the certification succeeds. Therefore, they are less desirable in
security-critical tasks (e.g., autonomous driving), since there remains a
non-trivial possibility that a ``certified'' NN violates
\E~\ref{equ:certification}, thus leading to severe outcomes and fatal errors.
\tool\ focuses on deterministic certification under semantic-level mutations,
although it is technically feasible to bridge \tool\ with probabilistic
certification frameworks as well.
\section{Generative Models and Image Mutations}
\label{sec:motivation}

Generative models like GAN~\cite{goodfellow2020generative},
VAE~\cite{kingma2014auto}, and GLOW~\cite{kingma2018glow} are popular for data
generation. As introduced in \S~\ref{subsec:transformation}, they are widely
used for approximating manifold, which encodes semantics of real-world images.
In brief, a generative model maps a collection of real-world images as points in
a low-dimensional latent space that follows a continuous distribution\footnote{
  The ``continuous'' indicates that coordinates in the latent space
  are continuous floating numbers, which does \textit{not} ensure the
  continuity in \S~\ref{sec:approach}.}
(e.g., normal distribution)~\cite{bengio2013representation,goodfellow2020generative}.
This way, ``intermediate images'' between two known images can be inferred.
Therefore, by randomly sampling points from the latent space, infinite diverse,
high-quality images can be generated.

Let the latent space of generative model $G$ be $\mathcal{Z}$,
generating an image $x$ can be expressed as $G: z \in \mathcal{Z} \rightarrow x
\in \mathcal{M}$, where $\mathcal{M}$ denotes the manifold of images. Note that
$\mathcal{M}$, approximated by $G(\mathcal{Z})$, does not equal the pixel
space of images, because most ``images'' with random pixel values are
meaningless. Conceptually, $\mathcal{M}$ is much smaller than the pixel
space.

\begin{figure}[!ht]
  \captionsetup{font=footnotesize}
  \centering
  \includegraphics[width=0.95\linewidth]{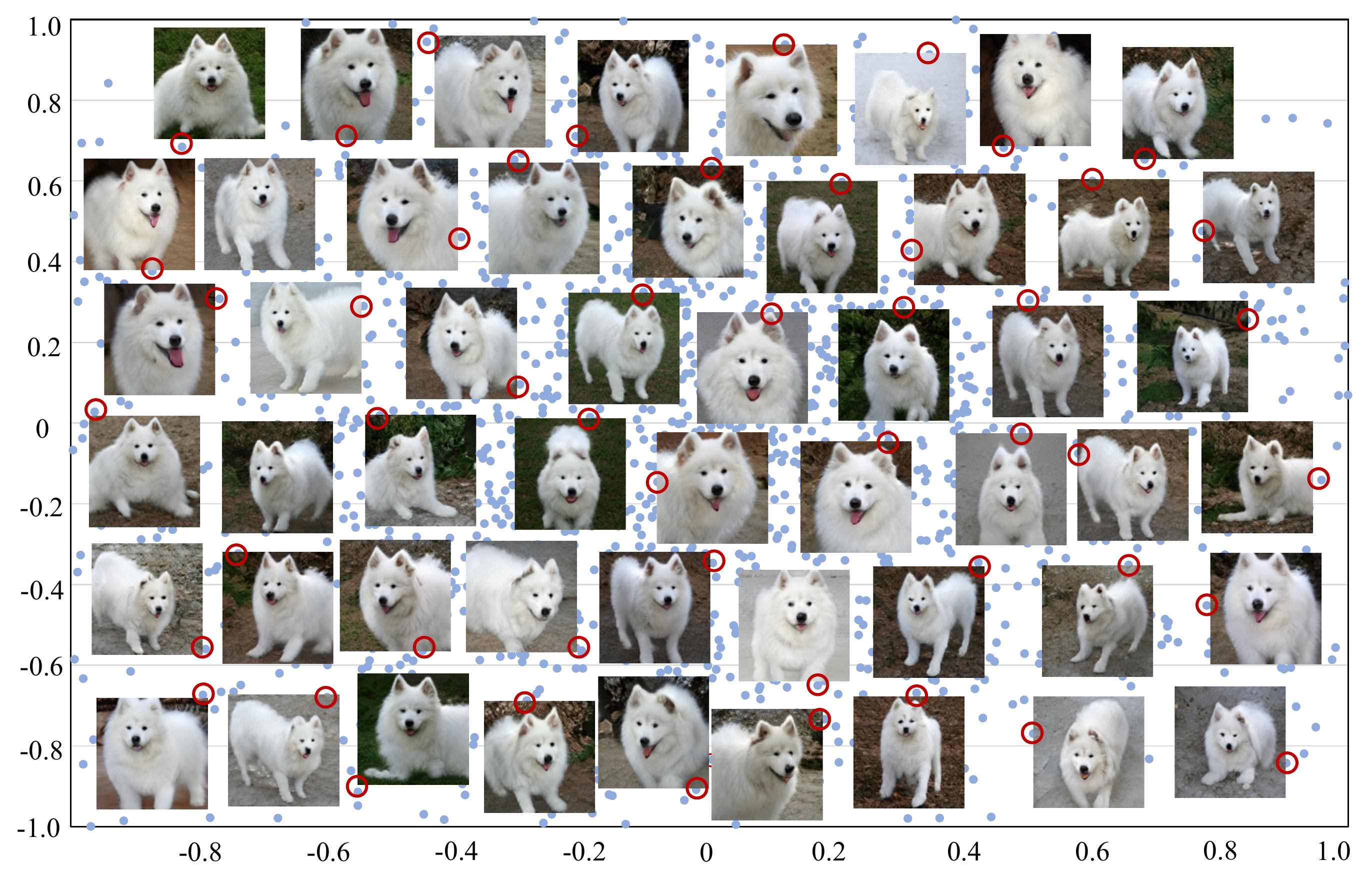}
  \vspace{-10pt}
  \caption{Moving $z$ in the latent space of BigGAN~\cite{brock2018large} and
  the derived images which have diverse semantics. Since the generative model is
  not regulated, these semantics change arbitrarily.}
  \label{fig:observation}
  \vspace{-5pt}
\end{figure}

\parh{Data-Driven Image Mutation.}~An important observation is that, given a
latent point $z$ corresponding to an image $x$, moving $z$ in the latent space
along certain directions results in images that differ from $x$ on the
semantics, as in \F~\ref{fig:observation}. Based on this observation,
\tool\ delivers semantic-level mutations (including both \mc\ and \ma\ listed
in \F~\ref{fig:mutation}) in a unified manner via generative models. 
We refer to this method as ``data-driven mutation'' since technically, the
latent space and the subsequently-enabled semantic-level mutations are deduced
from training images with diverse semantics.
Note that this is fundamentally \textit{distinct} from interpolation, an
image-processing tactic in GenProve~\cite{mirman2021robustness}. The
semantic-level mutations are inferred from training data with diverse
semantics, e.g., inferring a mutation ``opening/closing eyes'' from facial
images of \textit{different} human faces, as opposite to two images of the same
face with eyes opened/closed.

\parh{Data Preparation.}~It is essential to ensure meaningful semantics
are mutated. In fact, a semantic-level mutation is meaningful if it 
exists among real-life natural images. Since the newly enabled mutations are
``driven'' by semantics that vary in training data, we ensure that these
mutations are semantically meaningful by only using natural images as $G$'s
training data.

A general threat to the certifiable mutations is the out-of-distribution data of
$G$: \tool\ cannot certify mutations that are never seen by $G$ during training.
From an adversarial perspective, one may concern if attackers can leverage some
``never seen'' changes to create AEs. For instance, attackers leverage the
``open-eye'' mutation to create AEs for a facial recognition NN, but $G$ only
has training data of closed eyes. We clarify that this threat is out of the
consideration of \tool\ and all other certification tools in this field: it is
apparently that \tool\ (and all typical NN/software verification tools) will not
convey end-users a false sense of robustness towards unverified properties
(e.g., unseen mutations). On the other hand, as a typical offline certification
tool, \tool's main audiences are users who can access $G$. Thus, users may
fine-tune $G$ with concerned out-of-distribution data; it often takes marginal
efforts to turn them into in-distribution ones.

\parh{Two Requirements.}~We require each semantic-level mutation scheme should
correspond to a single direction in the latent space $\mathcal{Z}$. Thus, when
mutating $z$ along that direction, only the concerned property is
\textit{independently} mutated without changing other semantics (i.e.,
independence). This is intuitive and demanding: when checking the robustness
(fairness) of a facial recognition NN by mutating gender, we do not want to
change hair color. It also eases forensic analysis. If an auto-driving NN fails
certification, we know easily which property (e.g., road sign) is the root cause,
jeopardizing the NN.
Furthermore, when a latent point $z$ is moved in the latent space $\mathcal{Z}$,
the generated images should exhibit a \textit{continuous} change of semantics
(i.e., continuity). This allows the degree of mutation to be precisely controlled,
and ensure that latent points corresponding to all mutated images are exclusively
explored. Moreover, \S~\ref{sec:approach} clarifies that continuity
enables precise input space representation and complete certifications. 

\parh{Challenges.}~Generative models hardly provide an ``out-of-the-box''
solution for the two requirements. In \F~\ref{fig:observation}, we sample points
from the latent space of BigGAN~\cite{brock2018large}, a SOTA generative model,
to generate mutated images. For simplicity, we plot these points based on values
of their first two dimensions in the latent vectors, and show the accordingly
generated images in \F~\ref{fig:observation}. Clearly, when traversing along
each axis, several semantics (e.g., orientations, standing or not) that ought to
be \textit{independent}, are mutated together. Also, semantic changes are
non-continuous; there are \textit{sharp} changes on the semantics between two
images whose latent points stay close. See \S~\ref{sec:approach} for our technical solutions.
\section{Overview}
\label{sec:overview}

\begin{figure*}[!ht]
  \captionsetup{font=footnotesize}
  \centering
  \includegraphics[width=0.80\linewidth]{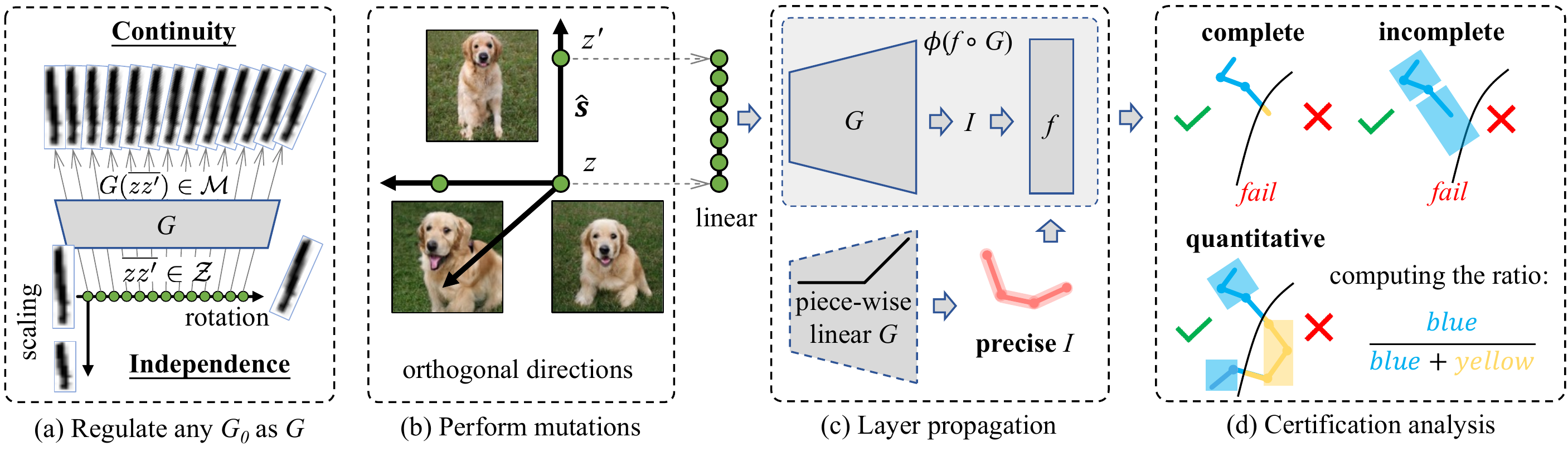}
  \vspace{-5pt}
  \caption{Overview of \tool. We illustrate the offline generative model
  regulation in (a), and the online input mutation and certification in
  (b)-(d).}
  \label{fig:overview}
\end{figure*}

\F~\ref{fig:overview} depicts the workflow of using \tool\ in NN certifications.
Below, we first discuss the offline and online phases in \F~\ref{fig:overview}.
We then analyze six key benefits of \tool\ in \cirC{1}--\cirC{6}.

\parhs{Offline Model Regulation:}~As in \F~\hyperref[fig:overview]{\ref*{fig:overview}(a)},
given a common generative model $G_o$, we
regulate $G_o$ with slight retraining. The ``regulation'' aims to satisfy two
requirements: independence and continuity, as noted in \S~\ref{sec:approach}
(see \S~\ref{subsec:independent} and \S~\ref{subsec:continuity} for the technical
details and correctness proof). The outcome of this phase is a regulated $G$,
such that when moving an image's latent point $z$ in $G$'s latent space, 1)
distinct semantics are mutated independently, and 2) the mutated semantics in
images $G(z)$ change continuously with $z$.

\parhs{Online Input Mutations:}~By mutating a latent point $z$ along various
orthogonal directions (due to independence) in the latent space of $G$,
different semantic-level mutations are achieved unifiedly (benefit
\cirC{1} and \cirC{2}; see below). Also, as shown in \F~\hyperref[fig:overview]{\ref*{fig:overview}(b)},
given a mutation and its corresponding direction $\hat{\bm{s}}$ (i.e., a unit vector)
in the latent space, latent points corresponding to all mutated inputs are exclusively
subsumed by a segment $\overline{zz'}$ (where $z' = z + \delta_{\max} \cdot \hat{\bm{v}}$)
in the latent space due to continuity. $G(z)$ is the original image and $G(z')$ is the
mutated image with the maximal extent.

\parhs{Online Layer Propagation:}~As in \F~\hyperref[fig:overview]{\ref*{fig:overview}(c)},
the segment $\overline{zz'}$ is propagated through $G$ to $f$ (our target NN) when conducting
certification.
Since $\overline{zz'}$ (i.e., the input space of $f \circ G$) is \textit{linear},
existing \textit{complete} certifications $\phi$ (which require linear input space) can be directly
applied on the new NN $f \circ G$ as $\phi(f \circ G(\overline{zz'}))$.
Moreover, when $G$ is piece-wise linear, a precise input space $I$ of $f$ can be
obtained with only \textit{polynomial} overhead (benefit \cirC{4}).
This precise input space, with different forms (e.g., per-pixel
lower/upper bound as in~\cite{balunovic2019certifying}), can be directly processed by 
existing certification frameworks.

\parhs{Certification Analysis:}~When the online certification propagation is
done, complete/incomplete (depending on if the layer propagation during
certification is complete or not) certification can be conducted to assess the
robustness of $f$. Since the input to $G$ is a segment, the output of $f$ will
be a \textit{chain of segments} (some ``segments'' may be over-approximated as
box/polyhedra\footnote{For activation functions that are \textit{not}
piece-wise linear.}~\cite{mirman2021robustness}). This representation enables
quantitative certification (see benefit \cirC{5}).

\parh{Requirement of $G_o$ and Training Data.}~\tool\ does not require a special
type of generative model $G_o$, and it converts $G_o$ into a regulated form $G$
automatically (see \S~\ref{sec:approach} and \S~\ref{sec:implementation}). We
evaluate \tool\ on common generative models in \S~\ref{subsec:eval-gen}. 
Also, $G$ has no special requirement on training data; we evaluate a set of
common training datasets in \S~\ref{subsec:eval-gen}. See
\S~\ref{sec:discussion} for further discussion on extensibility.

\parh{Key Benefits.}~\tool\ offers the following benefits:

\parhs{\cirC{1}~Advanced Mutation.}~\tool\ novelly enables
certifying NNs under advanced mutations listed in \F~\ref{fig:mutation}. These
mutations are rarely studied before, since they do not have explicit
math expression and are usually arbitrary: they can easily introduce
undesired perturbations which are hard to characterize. \S~\ref{sec:approach}
explains how these problems are overcome.

\parhs{\cirC{2}~Unified Formulation.}~Following \cirC{1}, \tool\ formulates
various semantic-level mutations in \F~\ref{fig:mutation} unifiedly:
each mutation is denoted using one direction in the latent space of 
$G$. Then, mutating an image $x$ is achieved by moving the
corresponding point $z$ along the direction in the latent space. The correctness
of this unified formulation is studied in \S~\ref{subsec:independent}. Compared
with prior certification frameworks considering one or a few specific
mutations, extending \tool\ to take into account new mutations is
straightforward; see \S~\ref{sec:discussion}.

\parhs{\cirC{3}~Practical Complete Certification over $f \circ G$.}~Since
existing approaches cannot form a precise input space representation for
semantic-level mutations, these mutations are never adopted by
prior complete certifications. We view this as a major, yet rarely discussed threat to NNs,
as semantic-level mutations generate images frequently
encountered in reality, and they
have been extensively studied in NN
testing~\cite{pei2017deepxplore,xie2019deephunter}. \tool\ empowers complete certification with
semantic-level mutations by applying complete certification on $f \circ G$ (see
results in \S~\ref{subsec:eval-app}). The key solution is to use linear segment
$\overline{zz'}$ as the input for certifying $f \circ G$. While it
appears that we add ``extra burden'' of certification by adding $G$, 
our enabled segment representation is however more efficient than prior ``region''
representation~\cite{xu2020fast,wang2018formal,
ferrari2021complete,wang2021beta}. Specifically, given a NN having $L$ layers
and $N$ neuron width (which is much larger than $L$), our strategy
reduces the cost from exponential
$O((2^{N})^{L})$~\cite{li2023sok} to $O((N^2)^L)$, which is \textit{polynomial}
when $L$ is fixed~\cite{sotoudeh2019computing};
see analysis in \Appx~\ref{appx:cost}.

\parhs{\cirC{4}~Precise Input Space $I$ over $f$.}~An important conclusion, as
noted by Sotoudeh et al.~\cite{sotoudeh2019computing}, is that given a
\textit{piece-wise linear} NN (e.g., ReLU as the activation function), if all of
its inputs can be exclusively represented by a segment, then all of its output
can also be exclusively represented as a chain of segments. This way, the input
space $I$ of $f$ can be precisely represented.
Following \cirC{3}, the cost of computing precise input space representation is
polynomial if $G$ has a fixed number of layers, which is true since $G$ does not
add layers during the certification stage. \S~\ref{subsubsec:bound} gives the
cost evaluation. Existing certification frameworks, by using our precise input
space (with different forms), can be smoothly extended to certify inputs mutated
by semantic-level mutations. \S~\ref{subsubsec:bound} compares previous
(imprecise) SOTA input space with ours. 

\parhs{\cirC{5}~Enabling Quantitative Certification.}\footnote{The same scheme
is referred to as probabilistic certification by Mirman et
al.~\cite{mirman2021robustness}. To distinguish it from the probability
guarantee (i.e., the NN is robustness with a high probability) derived from
certification works reviewed in \S~\ref{subsec:space}, we rename it in this
paper.} Complete certification can derive the maximum tolerance of NN $f$ to a
mutation~\cite{jordan2019provable,wang2021beta}. However, the NN may still
retain its correct prediction for some mutated inputs out of the maximum
tolerance, as shown in \F~\hyperref[fig:overview]{\ref*{fig:overview}(d)}. 
Quantitative certification provides fine-grained analysis by bounding the ratio
of mutated inputs whose derived predictions are
consistent~\cite{mirman2021robustness}. The key idea is to represent all NN
outputs as a chain of segments (see
\F~\hyperref[fig:overview]{\ref*{fig:overview}(d)}). Then, quantitative
certification is recast to computing the ratio of line segments that lie on the
proper side of the decision boundary. Recall as in \cirC{4}, the output of $f$ under
\tool-involved certification has already been a chain of line segments. Thus,
quantitative certification over various semantic-level mutations can be novelly
enabled. \S~\ref{subsubsec:eval-face} and
\S~\ref{subsubsec:eval-compare} evaluate quantitative
certification.

\parhs{\cirC{6}~Aligned Optimization in Incomplete Certification.}~In case
where cost is the main concern, incomplete certification is more preferred,
whose cost is $O(NL)$~\cite{li2023sok}. That said, certification (formal verification) is
inherently costly. One key objective in incomplete certification tasks is to
tune the trade-off between precision vs.~speed, by optimizing both input space
representation and layer propagations. Nevertheless, the two optimizations are
performed separately~\cite{balunovic2019certifying,li2023sok}, and in most cases, techniques in enhancing one
optimization cannot be leveraged to boost the other. In contrast, note that
\tool\ implements semantic-level mutations via $G$. Hence, incomplete
certification integrated with \tool, when optimizes layer propagation of $f
\circ G$, indeed covers both input space representation and layer propagation,
in a unified manner.

\section{Approach}
\label{sec:approach}

\parh{Problem Formulation.}~In accordance with semantic-level mutations listed
in \F~\ref{fig:mutation}, including all complex (\mc) and advanced (\ma)
mutations, our unified mutation is as follows.

\begin{definition}[Mutation]
    \label{def:mutation}
    Suppose $z \in \mathcal{Z}$ is the latent point of an image $x \in
    \mathcal{M}$\footnote{Mapping $x$ to $z$ is straightforward and
    well-studied~\cite{wang2022high}. Conventional generative models enable this
    mapping by design~\cite{kingma2014auto,kingma2018glow}.}, mutating $x$ as
    $x'$ using semantic-level mutations can be conducted via a generative model
    $G$
    \begin{equation}
    \small
    \label{equ:edit}
        x' = G(z + \| \delta \| \cdot \hat{\bm{s}}_{\tau}),
    \end{equation}
    \normalsize
    \noindent where $\hat{\bm{s}}_{\tau}$ is a unit vector (i.e.,
    $\norm{\hat{\bm{s}}_{\tau}} = 1$) that specifies the direction of one
    semantic-level mutation $\tau$. The norm of $\delta$, $\| \delta \|$,
    decides to what extent the mutation is performed.
\end{definition}

As in \S~\ref{sec:motivation}, to support NN certification, we require the
mutation in \Df~\ref{def:mutation} to satisfy the following two requirements.

\smallskip \noindent \underline{$\bullet$ \textbf{\texttt{Independence}}}:~When
moving $z$ in the latent space $\mathcal{Z}$ along $\hat{\bm{s}}_{\tau}$,
the resulting image should only change semantics associated to $\tau$. For
instance, mutating a portrait's eyes (e.g., by making them close) should not affect its
gender.

\smallskip \noindent \underline{$\bullet$ \textbf{\texttt{Continuity}}}:~Mutated
semantics must change continuously with $\| \delta \|$, such that points in
segment $\overline{zz'}$ that connects $z$ and $z' = z + \| \delta_{\max} \|
\cdot \hat{\bm{s}}_{\tau}$, when being used to generate images,
comprise exclusively all ``intermediate images'' between $G(z)$ and $G(z')$. For
instance, if $G(z')$ rotates $G(z)$ for $30^{\circ}$, $G(\overline{zz'})$
should include all rotated $G(z)$ within $30^{\circ}$ and no others.

\S~\ref{subsec:independent} and \S~\ref{subsec:continuity} show how
these two requirements are achieved. In \S~\ref{subsec:analysis}, we
prove how they ensure the soundness and completeness of certification.

\subsection{Achieving Independence}
\label{subsec:independent}

As an intuition, we first explore, when
moving $z$ in the latent space, what decides the mutated semantics.

\vspace{-5pt}
\begin{definition}[Taylor Expansion]
\label{def:expansion}
    Given a generative model $G$, for each $z$ and an infinitesimal $\Delta z$,
    according to Taylor's theorem, we can expand it as:
    \small
    \begin{equation}
    \label{equ:expansion}
    G(z + \Delta z) - G(z) = \mathbf{J}(z) \Delta z + O(\Delta z),
    \end{equation}
    \normalsize

    \noindent where $\mathbf{J}(z)$ is the local Jacobian matrix of $z$. Its
    $(i, j)$-th entry is $\mathbf{J}(z)_{i,j} = \frac{\partial G(z)_i}{\partial
    z_j}$, which is decided by $G$'s gradient calculated on $z$. $O(\Delta z)$
    approaches zero faster than $\Delta z$.
\end{definition}

\vspace{-5pt}
The expansion in \E~\ref{equ:expansion} suggests that for perturbations over
$z$, the changed semantics of the generated image are governed by the Jacobian matrix
$\mathbf{J}(z)$ over $z$. We then show how to decompose mutating directions
from $\mathbf{J}(z)$.

\vspace{-5pt}
\begin{theorem}[Degeneration~\cite{feng2021understanding}]
    \label{thm:degen}
    Let the dimensionality of $\mathcal{Z}$ be $d$ and $\mathbf{J}^{(k)}$
    be the Jacobian matrix of the input for the $k$-th layer of $G$.
    $\mathbf{J}^{(k)}$ is progressively degenerated with $k$:

    \small
    \begin{equation}
    \label{equ:degen}
        \mathrm{rank}({\mathbf{J}^{(k+1)}}^{\intercal} \mathbf{J}^{(k+1)})
        \leq \mathrm{rank}({\mathbf{J}^{(k)}}^{\intercal} \mathbf{J}^{(k)})
        \leq d
    \end{equation}
    \normalsize

    \noindent where $\mathrm{rank}(\cdot)$ denotes the rank of a matrix.
\end{theorem}

\Thm~\ref{thm:degen} suggests that,
when moving $z$ in $\mathcal{Z}$, not all dimensions of $\mathcal{Z}$ can
lead to semantic changes of $G(z)$, because the rank of
$\mathbf{J}^{\intercal}\mathbf{J}$ may decrease during the
forward propagation in $G$.\footnote{
    On the other hand, the ``degeneration'' is essential to enable
    local-perceptual mutations; as will be introduced in
    \S~\ref{subsec:unified}.} Thus, directly decomposing mutating directions
over $\mathbf{J}$ is inapplicable~\cite{zhu2021low}.
Therefore, to enable semantic-level mutations as in \Df~\ref{def:mutation},
we need to identify dimensions in $\mathbf{J}$ that correspond to effective
mutations. Following~\cite{zhu2021low}, we perform Low-Rank Factorization
(LRF) for the $\mathbf{J}^{\intercal}\mathbf{J}$ over $\mathcal{Z}$.

\parh{Solution.}~The LRF factorizes $\mathbf{J}^{\intercal}\mathbf{J}$ as
$\mathbf{J}^{\intercal}\mathbf{J} = \mathbf{R}^* + \mathbf{E}^*$, where
$\mathbf{R}^*$ is the low-rank matrix associated with effective semantics-level
mutations, and $\mathbf{E}^*$ is the noise matrix. The rank of $\mathbf{R}^*$,
$\mathrm{rank}(\mathbf{R}^*) = r \leq d$ where $d$ is the dimensions of
$\mathcal{Z}$, denotes the number of (independent) mutating directions (which is
affected by the training data of the generative model $G$; see \S~\ref{subsec:eval-gen}).
Thus, it becomes clear that to mutate each semantic independently, we can apply singular
value decomposition (SVD) for $\mathbf{R}^*$:
\small
\begin{equation}
\label{equ:svd}
    \mathbf{U}, \bm{\Sigma}, \mathbf{V}^{\intercal}
    = \mathrm{SVD}(\mathbf{R}^*),
\end{equation}
\normalsize

\noindent where $\mathbf{V} = [\hat{\bm{v}}_1, \dots, \hat{\bm{v}}_r,
\dots, \hat{\bm{v}}_d]$ is a concatenation of
\textit{orthogonal} vectors. Each $\hat{\bm{v}}$ is a unit vector corresponding
to one semantic-level mutation. Since different $\hat{\bm{v}}$ are
\textit{orthogonal}, performing one mutation will not move $z$ towards other
mutating directions. Therefore, we can
\textit{independently} mutate different semantics by setting $\hat{\bm{s}}$
in \E~\ref{equ:edit} as one vector from the first $r$-th vectors from
$\mathbf{V}$ (as displayed in \E~\ref{equ:edit2}). 
The remaining $d - r$ vectors in $\mathbf{V}$ do not lead to semantic
changes~\cite{zhu2021low}. These non-mutating directions enable independent
local mutations; see \S~\ref{subsec:unified} for details.
\small
\begin{equation}
\label{equ:edit2}
    x' = G(z + \| \delta \| \cdot \hat{\bm{v}}),
    \quad
    \hat{\bm{v}} \in [\hat{\bm{v}}_1, \dots, \hat{\bm{v}}_r].
\end{equation}
\normalsize

\subsection{Achieving Continuity}
\label{subsec:continuity}

As in \S~\ref{sec:motivation}, $G(\mathcal{Z})$ constructs the manifold
$\mathcal{M}$ which encodes semantics of images. Continuity requires the
constructed $\mathcal{M}$ to be continuous.
We define continuity in \Df~\ref{def:continuity} below.

\vspace{-5pt}
\begin{definition}[Continuity]
\label{def:continuity}
    Suppose the distance metrics on $\mathcal{Z}$ and $\mathcal{M}$ are
    $l_{\mathcal{Z}}$ and $l_{\mathcal{M}}$, respectively. 
    Then, $\forall z, z' \in \mathcal{Z}$, $\mathcal{M}$ is continuous iff
    it satisfies
    \small
    \begin{equation}
    \label{equ:continuous}
        \frac{1}{C} \, l_{\mathcal{Z}} (z, z')
        \; \leq \; l_{\mathcal{M}} (G(z), G(z'))
        \; \leq \; C \, l_{\mathcal{Z}} (z, z'),
    \end{equation}
    \normalsize
    \noindent where $C$ is a constant. The left bound ensures that different $z$
    on $\mathcal{Z}$ induce non-trivial changes of semantics. The right bound
    guarantees that the semantic changes continuously with $z$.
    $l_{\mathcal{Z}}$ can be computed as the Euclidean distance between two
    points in the latent space. $l_{\mathcal{M}}$ denotes the semantic-level
    similarity and is an implicit metric.
\end{definition}

\begin{figure}[!ht]
    \captionsetup{font=footnotesize}
    \vspace{-10pt}
    \centering
    \includegraphics[width=1.0\linewidth]{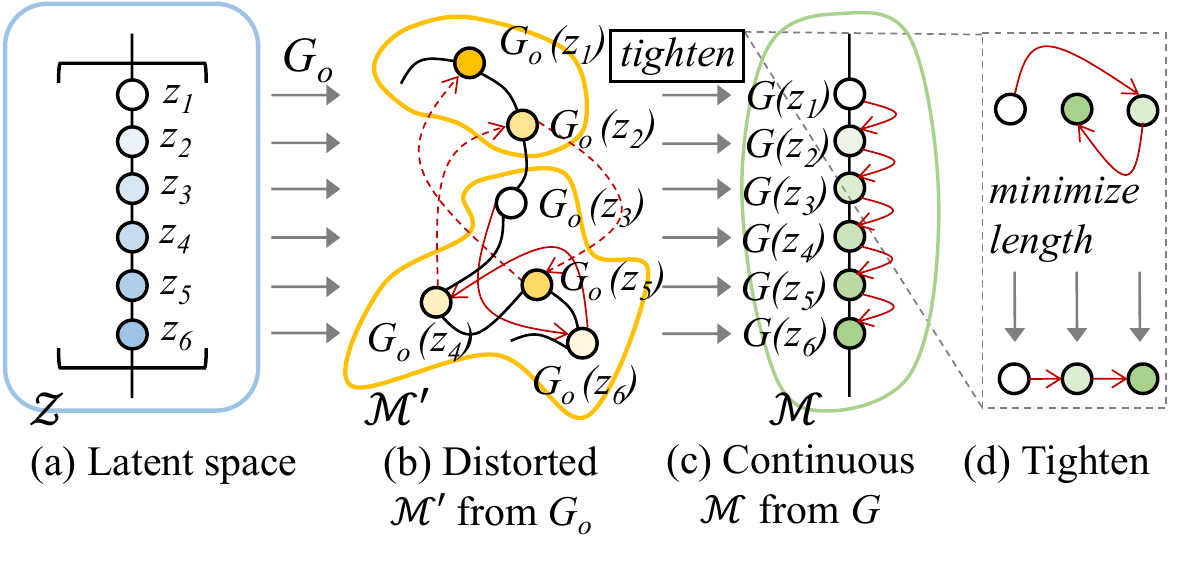}
    \vspace{-20pt}
    \caption{A schematic view of achieving continuity. The black lines ($-$) in
    (a)-(c) represent the intermediate points between two marked points. When
    mutating a point $z_i$, the red arrow (\textcolor{pptred}{$\rightarrow$}) in
    (b)-(d) marks the direction of changed $G_o(z_i)$ and $G(z_i)$.
    (a) depicts the latent space and the segment $\overline{z_1z_6}$ has
    intermediate points $z_2$-$z_5$. (b) shows a distorted $\mathcal{M}_o$,
    where $G_o(z)$ do not change continuously with $z$ and heavy distortion
    occurs. In (b), $G_o(z_3)$ is the minimally mutated one among all marked
    points, and the dashed red arrow (\textcolor{pptred}{$\dashrightarrow$})
    indicates an infinite length (as they cross two distorted sub-spaces scoped
    in yellow). (c) presents a continuous $\mathcal{M}$, where $G(z)$ yields
    gradual change with $z$, i.e., $G(z_1)$ and $G(z_6)$ denote the minimally
    and maximally mutated images, respectively. (d) illustrates the tightening
    procedure, which minimizes the length of red arrows
    (\textcolor{pptred}{$\rightarrow$}), thereby ensuring the continuity.} 
    \label{fig:smooth}
\end{figure}

\F~\hyperref[fig:smooth]{\ref*{fig:smooth}(b)} shows a non-continuous
$\mathcal{M}$ constructed by an ``out-of-the-box'' generative model $G_o$,
where heavy distortions occur on $G_o(z)$ when $z$ changes slightly. That is,
there are often \textit{sharp} changes on the semantics
between $G_o(z_i)$ and $G_o(z_{i+1})$ though $z_i$ and $z_{i+1}$ stay close. 
To make $\mathcal{M}$ continuous, our intuition is to ``tighten'' it: $\forall
z, z' \in \mathcal{Z}$, and for all latent points in $\overline{zz'}$, we
\textit{minimize the length of the curve} (which lies in $\mathcal{M}$) covered
by all points in $G(\overline{zz'})$. A schematic view of this tightening is in
\F~\hyperref[fig:smooth]{\ref*{fig:smooth}(d)}. The output would be a
regulated $G$, whose $\mathcal{M}$ is continuous, as in
\F~\hyperref[fig:smooth]{\ref*{fig:smooth}(c)}.
The following shows how a curve can be represented and how to
calculate curve length by integrating its speed.

\vspace{-2pt}
\begin{definition}[Curve Length]
    \label{def:curve}
        A curve connecting $a, a' \in \mathcal{A}$ is denoted as the trajectory
        of a point when it moves from $a$ to $a'$. It is formally represented as
        a mapping from a time interval to the space $\mathcal{A}$, along which
        the point moves:
        \small
        \begin{equation}
            \gamma_{\mathcal{A}}(t): [0, t] \rightarrow \mathcal{A},
        \end{equation}
        \normalsize

        \noindent where $\gamma_{\mathcal{A}}(t=0) = a$ and
        $\gamma_{\mathcal{A}}(t=T) = \overline{aa'}$. Accordingly, the length
        of a curve can be calculated as:
        \small
        \begin{equation}
            \mathrm{Len}[\gamma_{\mathcal{A}}(T)]
            = \int_0^T \| \dot{\gamma}_{\mathcal{A}}(t) \| \dif{t},
        \end{equation}
        \normalsize
    
        \noindent where $\dot{\gamma}_{\mathcal{A}}(t)
        = \frac{\mathrm{d} \gamma_{\mathcal{A}}(t)}{\dif{t}}$
        is the point's velocity at time $t$.
\end{definition}
\vspace{-2pt}

\parh{Problem Recast.}~Let $z_t$ be $z$'s location on $\mathcal{Z}$ at time $t$.
Then, $\gamma_{\mathcal{Z}}(T) = \overline{z_{t=0}\,z_{t=T}}$ and
$\gamma_{\mathcal{M}}(T) = G(\overline{z_{t=0}\,z_{t=T}})
= G(\gamma_{\mathcal{Z}}(T))$. We can thus compute the length of
$\gamma_{\mathcal{M}}(T)$ as:
\small
\begin{equation}
    \label{equ:length1}
    \begin{aligned}
        \mathrm{Len}[\gamma_{\mathcal{M}}(T)]
        = \int_0^T \norm{
                \frac{\partial G \circ \gamma_{\mathcal{Z}}(t)}{\partial \gamma_{\mathcal{Z}}(t)} 
                \frac{\partial \gamma_{\mathcal{Z}} (t)}{\partial t}
        } \dif{t}.
    \end{aligned}
\end{equation}
\normalsize

\noindent Suppose $\Delta t = \frac{T}{N}$ be an infinitesimal and
$\gamma_{\mathcal{Z}}(\Delta t) = \Delta z$. Let $t_i = i \Delta t$, then,
the length in \E~\ref{equ:length1} is reformulated as:
\small
\begin{equation}
    \label{equ:length2}
    \begin{aligned}
        \mathrm{Len}[\gamma_{\mathcal{M}}(T)]
        = \sum_{i=0}^{N} \norm{
            \mathbf{J}(z_{t_i})
            \frac{\Delta z}{\Delta t}
        } \Delta t
        = \sum_{i=0}^{N} \norm{
            \mathbf{J}(z_{t_i}) \Delta z
        }.
    \end{aligned}
\end{equation}
\normalsize

\noindent According to the Taylor expansion in \Df~\ref{def:expansion}, we have
\small
\begin{equation}
\label{equ:taylor1}
\begin{aligned}
    \| G(z_{t_i} + \Delta z) - G(z_{t_i}) \|
    = \| \mathbf{J}(z_{t_i}) \Delta z \|,
\end{aligned}
\end{equation}
\normalsize

\noindent Therefore, $\gamma_{\mathcal{M}}(T)$'s length in \E~\ref{equ:length1}
can be calculated as
\small
\begin{equation}
    \label{equ:discrete}
    \begin{aligned}
        \mathrm{Len}[\gamma_{\mathcal{M}}(T)]
        = \sum_{i=0}^{N} \norm{
            G(z_{t_i} + \Delta z) - G(z_{t_i})
        },
    \end{aligned}
\end{equation}
\normalsize

\noindent which is determined by the norm of $\mathbf{J}$.

\smallskip \parh{Solution.}~The above recast for
$\mathrm{Len}[\gamma_{\mathcal{M}}(T)]$ shows that, we can minimize
$\| \mathbf{J}(z) \| > 0$ for each $z$ to make $\mathcal{M}$ continuous.
Meanwhile, we should align the
starting and end points on two curves in $\mathcal{Z}$ and $\mathcal{M}$,
i.e., $\gamma_{\mathcal{M}}(0) = G(\gamma_{\mathcal{Z}}(0))$
and $\gamma_{\mathcal{M}}(T) = G(\gamma_{\mathcal{Z}}(T))$. Inspired
by the optimization objective in~\cite{ramasinghe2021rethinking},
we first define a monotonic function $\eta(\cdot)$ that satisfies
$\eta(0) = 0$ and $\eta(T) = 1$.\footnote{For simplicity, we let
$\eta(x) = \frac{1}{T} x$.} Then, for each $t_i = i \Delta t$,
we add an extra regulation term as in the following \E~\ref{equ:objective}
during the training stage of $G$. Overall, this regulation minimizes
the distance between $G(z_T)$ and $G(z_0)$. Simultaneously, it forces
$G(z_{t_i})$ to stay close to $G(z_T)$ if $z_{t_i}$ is close to $z_T$,
and vice versa for $z_0$ and $G(z_0)$.
As noted, the regulation term is agnostic to how the generative model
is implemented or trained; it can be applied on any generative model.
\small
\begin{equation}
\label{equ:objective}
 \argmin\limits_{\theta} \| 
        \eta(t_i) G_{\theta} (z_T) - G_{\theta}(z_{t_i})
        - (1 - \eta(t_i))G_{\theta}(z_0)
    \|
\end{equation}
\normalsize

\E~\ref{equ:objective} guarantees the continuity defined in \E~\ref{equ:continuous};
we provide detailed proof in \Appx~\ref{appx:proof}.

\begin{algorithm}[!htbp]
\footnotesize
\caption{Regulating $G_o$ with continuity.}
\label{alg:continuity}
\SetKw{KwBy}{by}
\SetKwProg{Fn}{function}{:}{}
    Distribution of latent space: $\mathcal{N}$; \tcp{Normal or uniform.}
    
    \Fn{$Loss${($G_o$, $\mathcal{N}$, $\dots$)}}{
    \tcp{The original training loss function of $G_o$ which returns a loss value.}
    }
    \Fn{$Continuity${($G_o$, $\mathcal{N}$)}}{
    $z_0 \sim \mathcal{N}$; \, $z_T \sim \mathcal{N}$\;
    $\lambda \sim [0, 1]$; \, $z_{t_i} \gets z_0 + \lambda (z_T - z_0)$\;
    \KwRet $\| \lambda G_o(z_T) - G_o(z_{t_i}) - (1 - \lambda) G_o(z_0) \|$\;
    }
    \Fn{$Regulating${($G_o$, $\mathcal{N}$, $\dots$)}}{
    \For{$i\gets0$ \KwTo $\#epochs$ \KwBy $1$}{
        $L_1 \gets Loss(G_o, \mathcal{N}, \dots)$\;
        $L_2 \gets Continuity(G_o, \mathcal{N})$\;
        \tcp{Optimize $G$ by minimizing $L_1 + L_2$.}
    }
    \KwRet Regulated $G$\;
    }
\end{algorithm}

The procedure to regulate a generative model is in \A~\ref{alg:continuity},
where line 7 is the entry point. \E~\ref{equ:objective} serves as an extra loss
term, which is added with the original training loss for optimization during the
training stage (line 10). The monotonic function $\eta(\cdot)$ in
\E~\ref{equ:continuous} helps avoid sampling many points: in each training
iteration, we randomly sample two points $z_0$, $z_T$ (line 4) and one point
in-between as $z_{t_i} = z_0 + \lambda (z_T - z_0)$ (line 5). $\eta(t_i) =
\frac{t_i}{T} = \lambda$ helps $G(z_{t_i})$ stay in the shortest path connecting
$G(z_0)$ and $G(z_{T})$ (line 6).
Regulating generative models with continuity should \textit{not} harm
their generation capability, since the original training objective is kept 
(line 9) during regulating. Moreover, continuity aims to
``re-organize locations'' of latent points (shown in \F~\ref{fig:smooth}) and
the left bound in \E~\ref{equ:continuous} ensures different $G(z)$ are generated
with different $z$.

\subsection{Soundness and Completeness Analysis}
\label{subsec:analysis}

\begin{algorithm}[t]
    \footnotesize
    \caption{Certification with \tool.}
    \label{alg:certification}
    \SetKw{KwBy}{by}
    \SetKwProg{Fn}{function}{:}{}
        The target NN: $f$; \, An input image: $x$ \;
        The regulated generative model: $G \gets Regulating(G_o, \dots)$\;
        A certification framework: $\phi$\;
        \tcp{$\phi$ can be complete, incomplete, or quantitative.}
        $z \gets G^{-1}(x)$; \tcp{Get the corresponding $z$ of $x$.}
        \tcp{Then get one mutating direction $\hat{\bm{s}}$ from $\mathbf{J}(z)$.}
        $z' \gets \|\delta_{\max}\| \cdot \hat{\bm{s}}$;

        \If{\cirC{3} or \cirC{5} or \cirC{6}}{
            $out \gets \phi(f \circ G(\overline{zz'}))$;
            \tcp{Optimize $f \circ G$'s propagations.}
        }
        \ElseIf{\cirC{4}}{
            $I \gets G(\overline{zz'})$;
            \tcp{Precise $I$ if $G$ is piece-wise linear.}
            $out \gets \phi(f(I))$\;
            \tcp{Works for complete/incomplete/quantitative $\phi$.}
            
        }
        \tcp{Analyze $out$ according to $\phi$.}
\end{algorithm}

Certifying NN robustness with \tool\ is given in \A~\ref{alg:certification}.
We demonstrate how \tool, which provides $G$ (line 2; see \A~\ref{alg:continuity}
for function $Regulating$) and $\overline{zz'}$ (line 4-5),
is incorporated into conventional certification framework $\phi$ under
scenarios discussed in \cirC{3}-\cirC{6} of \S~\ref{sec:overview}.

Below, we analyze the soundness and completeness. Overall, as in line 7 and
line 10, the whole certification procedure with \tool\ can be represented as
$\phi(f \circ G(\overline{zz'}))$, where $\overline{zz'}$ includes the corresponding
latent points of all mutated inputs. To ease the presentation, we also use the term
``sound'' and ``complete'' for $\overline{zz'}$. $\overline{zz'}$ is sound if it
does not miss any mutated input; it is complete if it only includes mutated inputs.

\begin{lemma}
    \label{lem:input}
        A certification $\phi(f \circ G(\overline{zz'}))$ is sound iff both
        $\phi$ and $\overline{zz'}$ are sound; it is complete iff both
        $\phi$ and $\overline{zz'}$ are complete. 
\end{lemma}

\noindent \textit{Proof.}~For a sound (but incomplete) $\phi$, the layer
propagation in $f \circ G$ is progressively over-approximated (e.g., via
abstract interpolation-based certification $\phi$~\cite{gehr2018ai2}). Since any
mutated inputs will not be missed during the propagation, $\phi(f \circ
G(\overline{zz'}))$ is sound if $\overline{zz'}$ is sound. For a complete (and
sound) $\phi$, the layer propagation in $f \circ G$ is precisely captured (e.g.,
via symbolic execution-based certification $\phi$~\cite{wang2018formal}).
Because the propagation does not introduce any new input, $\phi(f \circ
G(\overline{zz'}))$ is complete if $\overline{zz'}$ is complete.

Since we use conventional sound and/or complete $\phi$, based on \Lem~\ref{lem:input},
the soundness and completeness of $\phi(f \circ G(\overline{zz'}))$ is only decided
by $\overline{zz'}$. In the following, we analyze the soundness and completeness
of $\overline{zz'}$.

\parh{Soundness.}~The soundness of $\overline{zz'}$ is ensured by the continuity:
if $I$ misses some mutated inputs, there must exist two different $z_1$ and $z_2$ satisfying
$G(z_1) = G(z_2)$, which violates continuity as defined in \E~\ref{equ:continuous}.

\parh{Completeness.}~The completeness of $\overline{zz'}$ is ensured by both continuity
and independence. With independence, different mutations are represented as \textit{orthogonal}
directions in the latent space. This way, performing one mutation will not move the latent point
towards other unrelated directions. Also, when doing local mutations within one region, the mutating
direction is projected into the non-mutating direction of other unrelated regions (see \E~\ref{equ:proj}), 
such that only semantics over the specified region are mutated.
For continuity, if an undesirable input (which is not generated by the mutation) appear in
$\overline{zz'}$, for the corresponding latent point $z^*$, there exists $z^* + \epsilon$ (where $\epsilon \rightarrow 0$ is an infinitesimal),
such that $\| G(z^*) - G(z^* + \epsilon) \| > C \epsilon$, which contradicts the definition
in \E~\ref{equ:continuous}.

\subsection{Unified Semantic-Level Mutations}
\label{subsec:unified}

This section shows how semantic-level mutations, including geometrical,
stylized, global perceptual, and local perceptual mutations, are implemented
uniformly.

\subsubsection{Local Perceptual Mutations}

To enable local perceptual mutations for one specific
region $F$ (e.g., the eyes in \F~\ref{fig:mutation}), our intuition is to
identify a direction in $\mathcal{Z}$ towards which $F$ can be mutated, whereas
other regions are unchanged. Recall that in \S~\ref{subsec:independent}, LRF produces
a set of non-mutating vectors which do \textit{not} lead to
mutations (i.e., $[\hat{\bm{v}}_{r+1}, \dots, \hat{\bm{v}}_d]$ derived from
\E~\ref{equ:svd}). Thus, local perceptual mutations over $F$ can be performed by first
project $F$'s mutating directions into the non-mutating vectors of the remaining
region. To ease the presentation, let's divide an image as foreground, which is
the mutated region $F$, and background, which includes the remaining unchanged
regions. Let $\mathbb{P}_{x}$ be the set of all indices of $x$, then we have
\small
\begin{equation}
    \mathbb{P}_{x} = \mathbb{F}_{x} \cup \mathbb{B}_{x},
    \quad \emptyset = \mathbb{F}_{x} \cap \mathbb{B}_{x},
\end{equation}
\normalsize

\noindent where $\mathbb{F}_{x}$ and $\mathbb{B}_{x}$ denote two sets of indices
belonging to $F$ and background, respectively.
As in \E~\ref{equ:foreground}, we first calculate the Jacobian matrix
of $\mathbb{F}_{x}$:
\small
\begin{equation}
    \label{equ:foreground}
    \mathbf{J}^{\mathbb{F}}(z):
    \quad \mathbf{J}^{\mathbb{F}}(z)_{(i, j)} =
    \frac{\partial G(z)_i}{\partial z_j},
    \; i \in \mathbb{F}_{x}.
\end{equation}
\normalsize

\noindent Following our solution in \S~\ref{subsec:independent}, we apply
LRF and SVD on $\mathbf{J}^{\mathbb{F}}(z)$ to get
the corresponding $\mathbf{V}^{\mathbb{F}}$, whose rank is $r_{\mathbb{F}}$.
Then, by using $\mathbf{V}^{\mathbb{F}}$ as in \E~\ref{equ:edit2}, a total of
$r_{\mathbb{F}}$ mutating directions for the foreground is obtained.
Similarly, we repeat the same procedure for $\mathbb{B}_{x}$ and obtain
$\mathbf{V}^{\mathbb{B}} = [\hat{\bm{v}}_1, \dots,
\hat{\bm{v}}_{r_\mathbb{B}}, \dots, \hat{\bm{v}}_d]$, where
the last $d - r_\mathbb{B}$ vectors do not lead to mutations over
$\mathbb{B}$, i.e., the non-mutating vectors of the background.

Let $\mathbf{F} = [\hat{\bm{v}}_{1_{\mathbb{F}}}, \dots,
\hat{\bm{v}}_{r_{\mathbb{F}}}] \subset \mathbf{V}^{\mathbb{F}}$, and
$\mathbf{B} = [\hat{\bm{v}}_{r_{\mathbb{B}}+1}, \dots,
\hat{\bm{v}}_{d_{\mathbb{B}}}]$. Then, before mutating foreground
perceptions, we project each vector in $\mathbf{F}$ into $\mathbf{B}$
(see \E~\ref{equ:proj}). This way, only perceptions over the region
specified by $F$ are mutated.
\small
\begin{equation}
    \label{equ:proj}
    x' = G(z + \lambda \hat{\bm{s}}),
    \quad \hat{\bm{s}} = \mathbf{B}\mathbf{B}^{\intercal}\hat{\bm{v}},
    \quad \hat{\bm{v}} \in \mathbf{F}.
\end{equation}
\normalsize

\subsubsection{Geometrical and Global Perceptual Mutations}

Geometrical mutations can be
viewed as a special case of global perceptual mutations. Here, we first
perform ``data augmentation,'' by randomly applying various
geometrical mutations with different extents $\delta$ on images in the
training dataset. Our generative model $G$ is then trained with the augmented
training dataset. Note that we do \textit{not} need to explore all possible
$\delta$; the generative model can infer intermediate $\delta$ ($0 \leq \|
\delta \| \leq \| \delta_{\max} \|$) because the constructed $\mathcal{M}$ is
continuous (\S~\ref{subsec:continuity}). We also give empirical analysis in
\Appx~\ref{appx:ablation}.

When performing global perceptual (or geometrical) mutations, the whole image is
treated as the ``foreground.'' Therefore, we can easily follow
\E~\ref{equ:edit2} to perform mutations and satisfy the independence and
continuity requirements.

\parh{Clarification.}~Perceptual mutations are \textit{not}
arbitrary. As discussed in \S~\ref{sec:motivation}, generative models enable
``data-driven'' mutations --- the perception changes, therefore, are naturally
constrained by diverse perceptions in real-world data (which are used to train
our generative model $G$). For example, an eye will not be mutated into a nose,
since such unrealistic combinations of perceptions do not exist in real data.
Empirical evidences are provided in \Appx~\ref{appx:ablation}.

\subsubsection{Stylized Mutations}
\label{subsubsec:style}

Stylized mutations are widely
implemented via generative models (e.g., CycleGAN~\cite{zhu2017unpaired,isola2017image}) in an independent
manner. In particular, one individual generative model corresponds to one style,
and therefore, we only need to use our solution in \S~\ref{subsec:continuity} to
ensure their continuity.
Implementations of stylized mutations can be depicted as the following procedures.

\parh{Data Preparation.}~The training data of these generative models are
collected from two ``style domains'' $\mathcal{A}$ and $\mathcal{B}$. For
example, $\mathcal{A}$ consists of images taken on sunny days, while
$\mathcal{B}$ contains images taken on rainy days. Similarly, $\mathcal{A}$ may
likewise consist of natural images, whereas $\mathcal{B}$ contains artistic
paintings. The images in $\mathcal{A}$ and $\mathcal{B}$ do \textit{not} need to
have identical content (e.g., taking photos for the same car under different
weather conditions is \textit{not} necessary). 

\parh{Training Strategy.}~The training procedure is based on \textit{domain
transfer} and \textit{cycle consistency}~\cite{zhu2017unpaired,isola2017image}.
In practice, two generators, $G_{\mathcal{A}\mathcal{B}}$ and
$G_{\mathcal{B}\mathcal{A}}$, are involved in the training stage.
$G_{\mathcal{A}\mathcal{B}}$ maps images from domain $\mathcal{A}$ to domain
$\mathcal{B}$, while $G_{\mathcal{B}\mathcal{A}}$ performs the opposite. The
cycle consistency, at the same time, requires $\forall a \in \mathcal{A}, a =
G_{\mathcal{B}\mathcal{A}}(G_{\mathcal{A}\mathcal{B}}(a))$, and vice versa. This
allows the two generators to comprehend the difference between two style
domains, enabling style transfer. It is worth noting that each $G$ takes an
extra input $\delta$ to control to what extent the style is transferred.

\parh{Mutation \& Certification.}~As explained before, each generator is only
capable of performing mutations for a single style, hence different stylized
mutations are naturally implemented in an independent way. Therefore, to
incorporate stylized mutations into NN certification, it is only necessary to
establish continuity. To do so, our regulation term in
\E~\ref{equ:objective} is applied on $\delta$, as expressed below:
\small
\begin{equation*}
    \argmin\limits_{\theta} \| 
        \eta(t_i) G_{\theta} (a, \delta_T) - G_{\theta}(a, \delta_{t_i})
        - (1 - \eta(t_i))G_{\theta}(\delta_0)
    \|,
\end{equation*}
\normalsize

\noindent where $\delta_{t_i}$ increases with $t_i$.
The same applies for $G_{\mathcal{B}\mathcal{A}}$.

\section{Implementation}
\label{sec:implementation}

We implement \tool\ in line with \S~\ref{sec:approach}, including 1) the
solutions for the two requirements, 2) various semantic-level mutations, 3) the
support for different generative models, and 4) some glue code
for various certification frameworks, with a total of around 2,000 LOC in
Python. We ran experiments on Intel Xeon CPU E5-2683 with 256GB RAM and a Nvidia
GeForce RTX 2080 GPU.

\parh{Piece-Wise Linear.}~As explained in \S~\ref{sec:approach}, a piece-wise
linear $G$ is required to obtain precise input space $I$ for $f$ with a polynomial
cost. \tool\ is however not limited to piece-wise linear $G$. For instance,
when performing incomplete (but sound) certifications (which do not require a
precise $I$), $G$ can be any generative model after being regulated; existing
certification frameworks can be directly applied on $f \circ G$.
Note that the last layer of $G$ needs to convert its inputs from $x \in [-\infty,
\infty]$ into $[0, 1]$ or $[-1, 1]$ to generate images of valid pixel values
(as pixel values are rescaled into $[0, 1]$ or $[-1, 1]$~\cite{tfdata,torchdata}).
In practice, this layer is often implemented with
$Sigmoid(x): [-\infty, \infty] \rightarrow [0, 1]$ or
$Tanh(): [-\infty, \infty] \rightarrow [-1, 1]$ which are not piece-wise linear.
Therefore, we re-implement $G$'s last layer (which does not affect its capability;
see \S~\ref{subsec:eval-gen}) using the piece-wise linear function $ReLU(x) = \max(x, 0)$,
as $ReLU(-ReLU(x) + 1)$ for $[-\infty, \infty] \rightarrow [0, 1]$ and
$ReLU(-ReLU(x) + 2) - 1$ for $[-\infty, \infty] \rightarrow [-1, 1]$.

\section{Evaluation}
\label{sec:evaluation}

Evaluations of \tool-enabled mutations are launched in \S~\ref{subsec:eval-gen}.
Then, \S~\ref{subsec:eval-app} studies the input space representation of \tool\
and incorporates \tool\ into different certification frameworks for various
applications. We also summarize lessons we learned from the evaluations.

\subsection{Mutations and Generative Models}
\label{subsec:eval-gen}

\parh{Evaluation Goal.}~This section corresponds to \cirC{1} and \cirC{2}
discussed in \S~\ref{sec:overview}. It serves as the empirical evidence besides
the theoretical analysis of correctness in \S~\ref{sec:approach}.
We first launch qualitative and quantitative evaluations
to assess the correctness and diversity of various $G$-supported mutations.
Due to the limited space, ablation studies of how $G$'s capability is affected
are presented in \Appx~\ref{appx:ablation}.

\begin{table}[!htpb]
    \caption{Evaluated datasets and generative models.}
    \vspace{-5pt}
    \label{tab:data}
    \centering
	\setlength{\tabcolsep}{2.0pt}
  \resizebox{1.0\linewidth}{!}{
    \begin{tabular}{l|c|c|c|c|c}
      \hline
      \textbf{Dataset}   & MNIST~\cite{deng2012mnist} & CIFAR10~\cite{krizhevsky2009learning} & CelebA~\cite{liu2015faceattributes} & CelebA-HQ~\cite{karras2018progressive} & Driving~\cite{driving} \\
      \hline
      \textbf{Size} & 32 $\times$ 32 &  32 $\times$ 32 & 64 $\times$ 64 & 256 $\times$ 256 & 256 $\times$ 128 \\
      \hline
      \textbf{Task} & \multicolumn{2}{c|}{Classification} & \multicolumn{2}{c|}{Face recognition} & Auto driving \\
      \hline
      \textbf{Color} & Gray-scale & \multicolumn{4}{c}{RGB} \\
      \hline
      \textbf{$G$} & DCGAN*~\cite{radford2015unsupervised} & BigGAN~\cite{brock2018large} & WGAN*~\cite{arjovsky2017wasserstein} & StyleGAN~\cite{karras2019style} & CycleGAN~\cite{zhu2017unpaired} \\
      \hline
    \end{tabular}
    }
    \begin{tablenotes}
        \footnotesize
        \item * DCGAN/WGAN is configured as piece-wise linear following
        \S~\ref{sec:implementation}.
    \end{tablenotes}
\end{table}

\parh{Datasets.}~\T~\ref{tab:data} lists the used datasets.
MNIST consists of images of handwritten digits with a single
color channel. Images in other datasets are RGB images. CIFAR10
contains real-life images of ten categories and is typically adopted for
classification. CelebA-HQ consists of human face photos of much larger size.
Note that CelebA-HQ images have \textit{exceeded} the capacity of contemporary
certification frameworks. Nevertheless, we use CelebA-HQ to demonstrate that more
fine-grained mutations can be enabled in \tool. Images of higher resolution
generally support more fine-grained mutations. Driving consists of driving-scene
images captured by a dashcam and is mostly used to predict steering angles using
NNs. In sum, these datasets are representative.

\parh{Generative Models.}~We employ popular generative models listed in
\T~\ref{tab:data} (see implementation details in~\cite{snapshot}).
As introduced in \S~\ref{sec:approach}, our enforcement of independence and
continuity is agnostic to the implementation of generative models. We assess
several commonly-used generative models to illustrate the generalization of our
approach.

\begin{figure}[!ht]
    \captionsetup{font=footnotesize}
    \centering
    \includegraphics[width=0.95\linewidth]{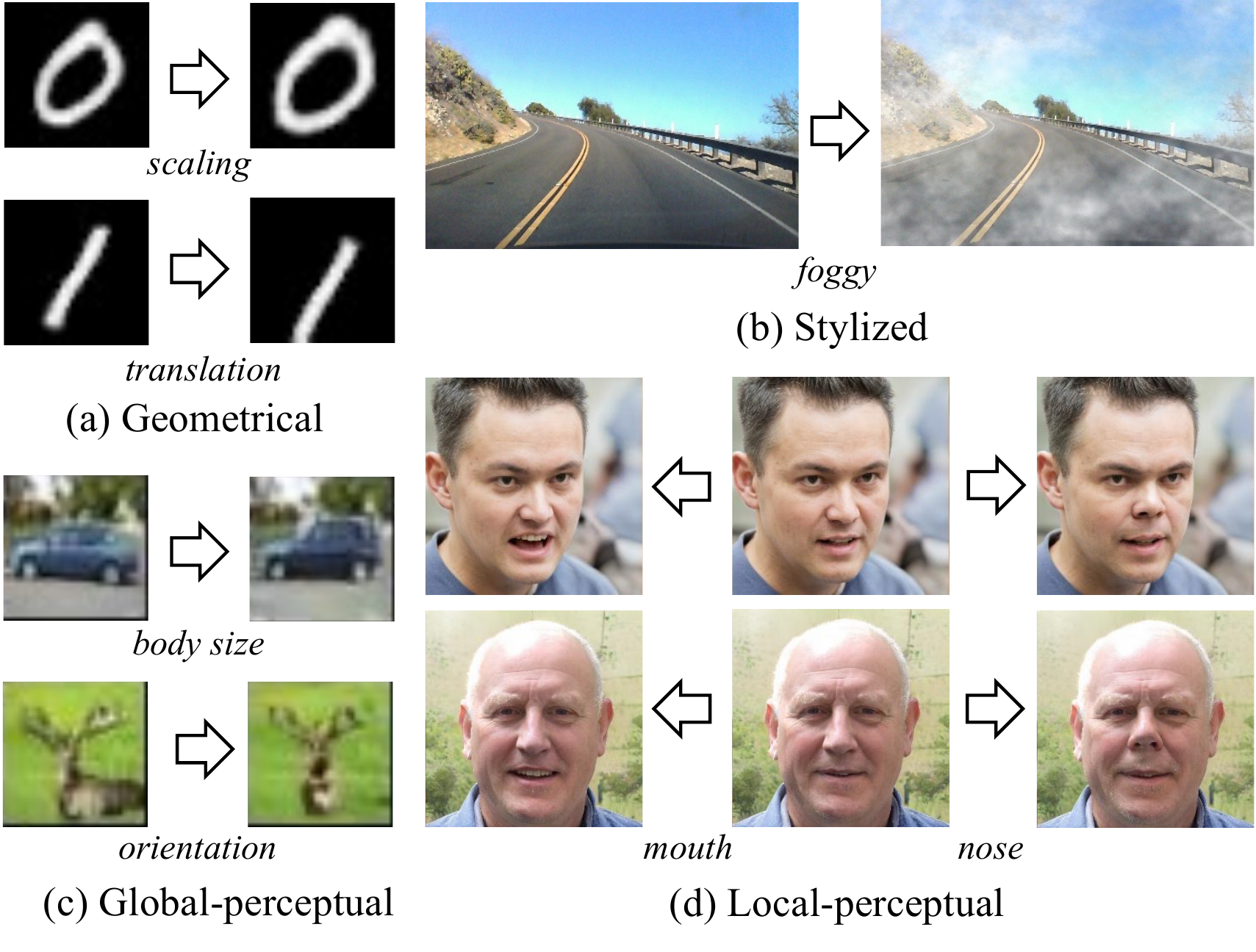}
    \vspace{-10pt}
    \caption{Independence evaluation. To clarify, CIFAR10 images in (c)
    generally have low resolution; \tool\ does not undermine its resolution.}
    \label{fig:eval-ind}
    \vspace{-10pt}
\end{figure}

\subsubsection{Qualitative Evaluation}
\label{subsubsec:qualitative}

\parh{Independence.}~\F~\ref{fig:eval-ind} shows images mutated using different
type of mutations listed in \F~\ref{fig:mutation}. Compared with the original
images, mutations enabled by \tool\ are \textit{independent}. For example, in
the lower case of \F~\hyperref[fig:eval-ind]{\ref*{fig:eval-ind}(a)}, the
translation scheme only changes the position of the digit, but retains the scale
and the rotating angle. In the upper case of
\F~\hyperref[fig:eval-ind]{\ref*{fig:eval-ind}(d)}, when mutating the mouth, other
facial attributes are unchanged.

\begin{figure}[!ht]
    \captionsetup{font=footnotesize}
    \centering
    \includegraphics[width=0.9\linewidth]{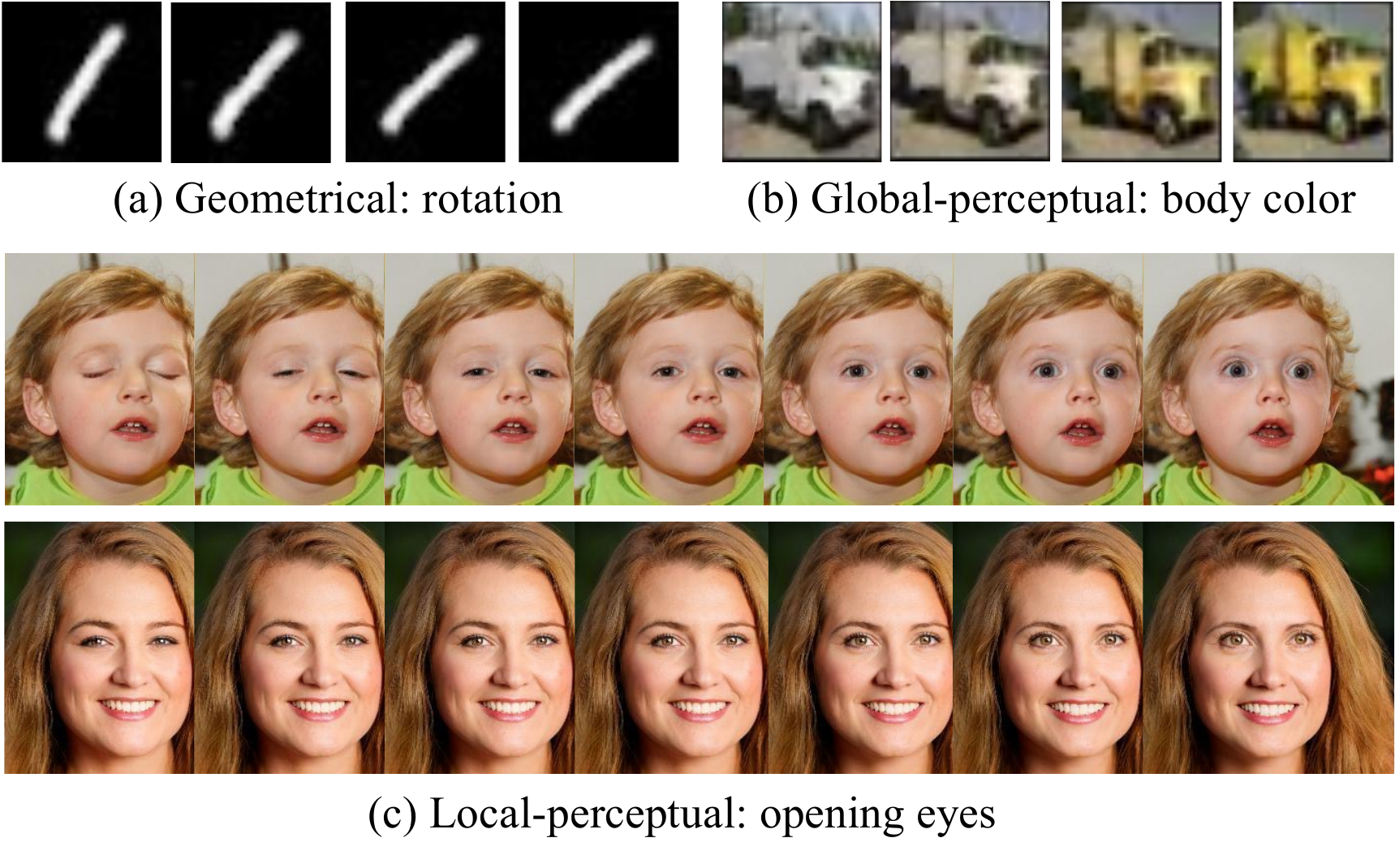}
    \vspace{-10pt}
    \caption{Continuity evaluation. To clarify, CIFAR10 images in (b) generally
    have low resolution; \tool\ does not undermine its resolution.}
    \label{fig:eval-con}
    \vspace{-10pt}
\end{figure}

\parh{Continuity.}~\F~\ref{fig:eval-con} presents a sequence of changed images
when gradually increasing $\delta$. Clearly, the mutated properties change
continuously with $\delta$. Moreover, from
\F~\hyperref[fig:eval-con]{\ref*{fig:eval-con}(a)}, it is apparent that the
rotation degree of the second image is bounded by the first (the original one)
and the third images, despite that the digit is only slightly rotated when
comparing the first and the third images. 

\subsubsection{Quantitative Evaluation}
\label{subsec:quant-eval}

\parh{Correctness.}~Since it is challenging to obtain ground truth for
stylized and perceptual mutations, to evaluate the correctness
of mutations achieved by $G$, we focus on geometrical mutations. Ideally,
we can first perform geometric mutations as ground truth, and then compare
them with results of $G$ in a pixel-to-pixel manner. However, as mentioned
in \S~\ref{subsec:transformation}, geometrical mutations perform interpolations
because pixel indexes are integers. Different interpolation algorithms will
produce images of different pixel values, but the geometrical mutations are
equivalent. Also, objects in images are irregular. As a result, obtaining
their geometric properties (e.g., size, rotating angle) is challenging.

\begin{figure}[!ht]
   \captionsetup{font=footnotesize}
    \centering
    \includegraphics[width=0.80\linewidth]{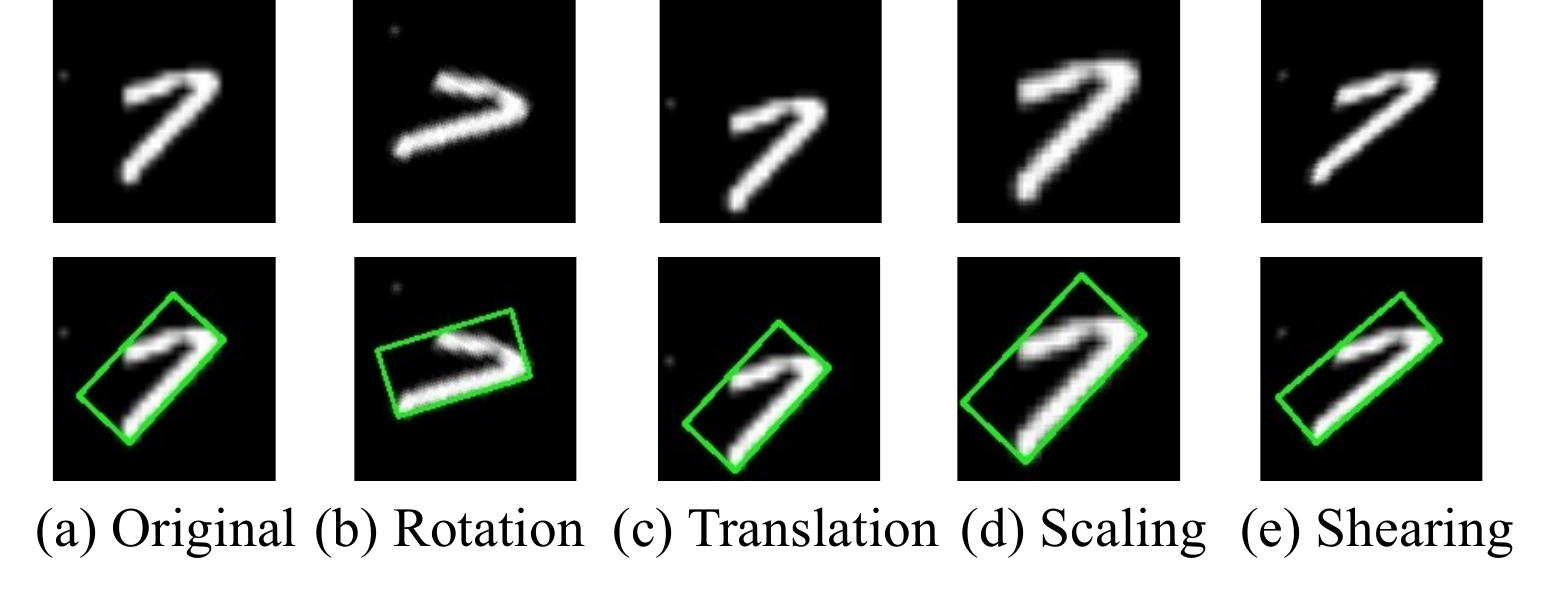}
    \vspace{-10pt}
    \caption{Minimum enclosing rectangles of objects.}
    \label{fig:eval-rect}
\end{figure}

Nevertheless, it is unnecessary to have the exact geometric properties, since
we only need to check whether the geometric properties change independently and
continuously. Therefore, we compute the minimum
enclosing rectangle of an object, as shown in \F~\ref{fig:eval-rect}. We then
check whether independence and continuity hold for geometric properties of this
rectangle when the image is gradually mutated. In particular, to evaluate
independence, we check whether other geometrical properties are unchanged when
gradually mutating one geometrical property. Since shearing mutation warps
the object horizontally or/and vertically, it will change the size and rotation
angle of enclosing rectangle. Thus, it is infeasible to check the independence
between shearing and scaling/rotation.\footnote{Shearing and scaling/rotation
are only ``dependent'' in the view from minimum enclosing rectangle. For the
mutated object, the distance between two parallel lines (in the original object)
is retained by shearing but changed by scaling; these two mutations are indeed
\textit{independent}.} As shown in \T~\ref{tab:eval-ind}, these mutations
satisfy the independence.

To evaluate continuity, for each geometrical property, we first \cirC{a}
randomly generate two images $x_1$ and $x_2$ (by mutating $x_1$ as $x_2$) whose
geometrical differences are $\Delta$ (e.g., the rotated angle is $30^{\circ}$).
Then, for the corresponding latent points $z_1$ and $z_2$, we \cirC{b} randomly
sample points $z'$ from the segment $\overline{z_1 z_2}$. For every sampled
point $z'$, we obtain $x' = G(z')$ and check whether the geometrical difference
between $x'$ and $x_1$ (or $x_2$) is no greater than $\Delta$. We repeat both
\cirC{a} and \cirC{b} for 100 times which results in total 10,000 checks. We
checked $\Delta$ of different scales (i.e., $\Delta_1$ and $\Delta_2$ in
\T~\ref{tab:eval-con}). The passed check ratios for different geometrical
mutations are reported in \T~\ref{tab:eval-con}. Clearly, all points satisfy the
continuity requirement.

\begin{table}[t]
    \caption{Quantitative evaluation for independence.}
    \vspace{-10pt}
    \label{tab:eval-ind}
    \centering
  \resizebox{0.8\linewidth}{!}{
    \begin{tabular}{c|c|c|c|c}
      \hline
            & Translation & Rotation & Scaling & Shearing \\
      \hline
        Translation & N/A & \cBrush & \cBrush & \cBrush \\
      \hline
        Rotation & \cBrush & N/A & \cBrush & N/A \\
      \hline
        Scaling & \cBrush & \cBrush & N/A & N/A \\
      \hline
        Shearing & \cBrush & N/A & N/A & N/A \\
      \hline
    \end{tabular}
    }
    \begin{tablenotes}
        \footnotesize
        \item* \cBrush\ means the
    mutation in the row does not influence the mutation in the column header.
    \end{tablenotes}
   \vspace{-5pt}
\end{table}

\begin{table}[!htpb]
    \caption{Quantitative evaluation for continuity*.}
    \vspace{-5pt}
    \label{tab:eval-con}
    \centering
  \resizebox{0.7\linewidth}{!}{
    \begin{tabular}{c|c|c|c|c}
      \hline
         & Translation & Rotation & Scaling & Shearing \\
      \hline
        $\Delta_1$ & 100\% & 100\% & 100\% & 100\% \\
      \hline
        $\Delta_2$ & 100\% & 100\% & 100\% & 100\% \\
      \hline
    \end{tabular}
    }
    \begin{tablenotes}
        \footnotesize
        \item * $\langle \Delta_1, \Delta_2 \rangle$ are
        $\langle \pm 10, \pm 4 \rangle$, $\langle \pm 30^{\circ}, \pm 10^{\circ} \rangle$,
        $\langle \pm 50\%, \pm 20\% \rangle$, $\langle \pm 10, \pm 4 \rangle$
        for translation, rotation, scaling, and shearing, respectively.
    \end{tablenotes}
   \vspace{-5pt}
\end{table}

\parh{Diversity.}~Geometric mutations are of a fixed number, and each stylized
mutation is separately implemented (i.e., one $G$ supports one style).
Hence, we focus on evaluating the diversity of perceptual mutations.
As discussed in \S~\ref{subsec:independent}, after performing LRF for the Jacobian
matrix $\mathbf{J}$, the rank of the obtained low-rank matrix decides the number of
independent mutations. For the CelebA-HQ dataset, the rank is
around 35. Clearly, a considerable amount of perceptual mutations are enabled;
see \Appx~\ref{appx:ablation} for results of CIFAR10 ranks.
We present more cases of perceptual mutations in our artifact~\cite{snapshot}.

\subsection{Certification and Applications}
\label{subsec:eval-app}

\parh{Evaluation Goal.}~This section first collects the precise input space for
various semantic-level mutations offered by \tool, and compares them with bounds
yielded by previous approaches in \S~\ref{subsubsec:bound}
(i.e., \cirC{4} in \S~\ref{sec:overview}).
Then, using different certification frameworks, we certify various NNs with
semantic-level mutations in different scenarios.

\S~\ref{subsubsec:eval-classification} leverages \tool\ to perform complete
certification and studies how NN architectures and training
strategies affect its maximal tolerance to different mutations.
\S~\ref{subsubsec:eval-face} incorporates \tool\ into existing quantitative
certification frameworks to certify perceptual-level mutations over face
recognition. \S~\ref{subsubsec:eval-drive} conducts incomplete certification
(which has the highest scalability) with \tool\ for autonomous driving.
\S~\ref{subsubsec:eval-compare} systematically evaluates NN robustness towards
all different types of mutations.

\begin{table}[!htpb]
    \caption{Comparing average/median distances between lower and upper bounds yielded
    by \tool\ and DeepG.}
    \vspace{-5pt}
    \label{tab:eval-bound}
    \centering
  \resizebox{0.95\linewidth}{!}{
    \begin{tabular}{c|c|c|c|c|c}
      \hline
        \multirow{2}{*}{Tool}  & \multicolumn{2}{c|}{Rotation $\pm 30^{\circ}$} & \multicolumn{2}{c|}{Translation $\pm 4$} & \multirow{2}{*}{Precise$^*$} \\
      \cline{2-5}
       & Avg./Med. Distance & Speed & Avg./Med. Distance & Speed & \\
      \hline
        DeepG  & 0.120/0.116 & 2.12s & 0.144/0.146 & 2.57s & {\xBrush} \\
      \hline
        \tool\  & \textbf{0.075}/\textbf{0.076} & \textbf{0.16s} & \textbf{0.087}/\textbf{0.083} & \textbf{0.21s} & \cBrush \\
      \hline
    \end{tabular}
    }
    \begin{tablenotes}
      \footnotesize
      \item * Whether the tool delivers a precise input space.
  \end{tablenotes}
   \vspace{-15pt}
\end{table}

\subsubsection{Precise Input Space}
\label{subsubsec:bound}

We compare \tool\ with the SOTA approach DeepG~\cite{balunovic2019certifying}, a deterministic certification
tool. As noted in \S~\ref{subsec:space}, probabilistic approaches determine if
the NN is robust with a probability. We deem this as undesirable, and thus
omit comparison with probabilistic approaches. To offer a precise input
space representation, \tool\ uses a piece-wise linear $G$ (DCGAN in
\T~\ref{tab:data}) at this step. In contrast, SOTA approaches like DeepG
over-approximate the input space. 

\parh{Setup.}~Following DeepG, we first randomly select 100 images from the
MNIST dataset. Then, for each image, we compute the \textit{precise} input space
resulted from rotation/translation (we follow the same setting as DeepG and
configurations are listed in \T~\ref{tab:eval-bound}) using \tool, and obtain
the upper/lower bounds for every pixel value. Finally, we compute the
average distance between the upper and lower bounds, and compare the results
with that of DeepG. 
To clarify, the precise input space representation yielded by \tool\ is
\textit{not} a set of lower/upper bounds. Instead, it is a chain of segments
generated by the piece-wise linear $G$ in \tool. To use precise input space
for certification, users directly pass the chain into the target NN for
certification. Here, we compute the lower/upper-bound form over our generated
chain to ease the comparison with DeepG, as it computes lower/upper bounds.

\parh{Results.}~As presented in \T~\ref{tab:eval-bound}. \tool\ reduces the
average/median distances by around $40\%$
compared with DeepG. In addition, \tool's speed should not be a concern. As reported
in \T~\ref{tab:eval-bound}, \tool\ takes about 0.2s to analyze one image,
compared to 2s for DeepG. Training $G$ --- a one-time effort for all images
--- takes about 150s. However, we do \textit{not} take credit from speed. We
clarify that DeepG computes bounds for each pixel on CPUs \textit{in parallel}
while \tool\ executes on a GPU (i.e., a forward pass of $G$).

We also benchmark probabilistic certifications; we use TSS~\cite{li2021tss},
the SOTA framework. Probabilistic certifications use the Monte Carlo algorithm
to sample mutated inputs when estimating the input space, which is \textit{not}
aligned with the input space in deterministic certifications. Nevertheless, we
report the time cost of sampling (on the same hardware with \tool) for reference. In practice,
the time cost is mainly decided by the number of samples, for example, sampling
100K mutated inputs (following the default setup) takes 5--6s and only 0.05s 
is need if sampling 1K inputs. We interpret its speed is reasonable and 
comparable to that of \tool.

\begin{table}[!htpb]
    \caption{NNs and their tolerance to different mutations.}
    \vspace{-5pt}
    \label{tab:eval-class}
    \centering
  \resizebox{0.95\linewidth}{!}{
    \begin{tabular}{l|c|c|c|c|c|c|c|c}
        \hline
              & \#Conv$^{1}$ & \#FC$^{1}$ & \#Layer & Aug.$^{2}$ & Speed & Max. R$^{3}$ & Max. T$^{4}$ & Max. S$^{5}$\\
        \hline
        $f_1$ & 0 & 3 & 3 & \xBrush & 0.98s & $41.1^{\circ}$ & 6.5 & 32\% \\
        \hline
        $f_2$ & 2 & 1 & 3 & \xBrush & 1.76s & $52.9^{\circ}$ & 7.6 & 39\% \\
        \hline
        $f_3$ & 4 & 2 & 6 & \xBrush & 3.86s & $72.3^{\circ}$ & 9.1 & 44\% \\
        \hline
        $f_4$ & 4 & 2 & 6 & \cBrush & 3.85s & $90.0^{\circ}$ & 9.3 & 51\% \\
        \hline
    \end{tabular}
    }
    \begin{tablenotes}
        \footnotesize
        \item 1. Conv is convolutional layer and FC is fully-connected layer.
        \item 2. Aug. flags whether the NN is trained with data augmentation,
        where during training, geometrical mutations are applied on the training
        data to increase the amount and diversity of training samples.
        \item 3. Max. R/T/S denotes the averaged maximal tolerance under rotation/translation/scaling
        where the NN consistently retains its prediction.
        \item 4. The numerical unit is pixel.
        \item 5. The tolerance $x\%$ denotes scaling within $[1 - x\%, 1 + x\%]$.
    \end{tablenotes}
   \vspace{-10pt}
\end{table}

\subsubsection{Complete Certification for Classification}
\label{subsubsec:eval-classification}

\parh{Setup.}~We perform complete certification for NNs listed in
\T~\ref{tab:eval-class}. These NNs are composed of different numbers/types
of layers and trained with different strategies. They perform 10-class
classification on the MNIST dataset. We certify their robustness using 100 images
randomly selected from the test dataset of MNIST. \tool\ offers complete
certification, meaning if the certification fails, the NN must have mis-predictions.
For each image, we record the maximal tolerance (within which the NN retains
its prediction) to different mutations. For each NN, we report the average
tolerance over all tested images in \T~\ref{tab:eval-class}. A higher tolerance
indicates that the NN is more robust to the mutation.
Since we require a meaningful metric to measure the tolerance, we focus on
geometrical mutations which have explicit math expressions: the tolerance
can be expressed using interpretable numbers.
In this evaluation, \tool\ is bridged with ExactLine~\cite{sotoudeh2019computing},
a complete certification framework.

\parh{Conv vs. FC.}~As shown in \T~\ref{tab:eval-class},
these NNs manifest different degrees of robustness towards various mutations.
Comparing $f_1$ and $f_2$, we find that replacing FC with Conv layers can
enhance the robustness. It is known that Conv, due to its sparsity, is more robust
if the input image is seen from different angles (which is conceptually equivalent
to rotation/translation/scaling)~\cite{lecun2004learning,zeiler2014visualizing}.
Our results support this heuristic.

\parh{Depth.}~Comparing $f_2$ and $f_3$ reveals that increasing the depth
(adding more layers) of NNs also enhances the robustness. According to prior
research~\cite{schmidhuber2015deep}, deeper layers in
NNs extract higher-level concepts from inputs, e.g., ``digit 1'' is a
higher-level concept than ``digit position.'' Consequently, deeper NNs should be
more resilient to perturbations that change low-level image concepts (which
correspond to the primary effect of geometrical mutations). Our results justify
this hypothesis from the certification perspective.

\parh{Data Augmentation.}~The tolerance increase of $f_4$ relative to $f_3$
shows the utility of data augmentation (by randomly performing geometrical
mutations on training data) for enhancing robustness. Applying geometrical
mutations on training data ``teaches'' the NN to not concentrate on specific
geometrical properties (since these properties may have been diminished in the
mutated training data). Previous works mostly use data augmentation to increase
the accuracy of NNs since more data are used for training. Our evaluation offers
a new viewpoint on the value of data augmentation.

\parh{Speed.}~\T~\ref{tab:eval-class} shows the time of analyzing one image
in \tool. Though complete certification was believed costly, it is
relatively fast in \tool: for the largest NNs ($f_3$
and $f_4$), one complete certification takes less than 4s. The main reason is
that in previous complete certification, all inputs are represented as a region,
whereas they are simplified as a segment to $f \circ G$ in \tool. We give
detailed complexity analysis in \Appx~\ref{appx:cost}.

\subsubsection{Quantitative Certification for Face Recognition}
\label{subsubsec:eval-face}

Besides incorporating \tool\ with complete certification, we use \tool\ for
quantitative certification below.

\parh{Setup \& Framework.}~We incorporate \tool\ into
GenProver~\cite{mirman2021robustness} for quantitative certification: it
delivers more fine-grained analysis by computing the lower/upper bound for the
percentage of mutated inputs that retain the prediction. The target NN takes
two face photos as inputs and predicts whether they belong to the same person
(i.e., predicts a number in $[0, 1]$ to indicate the probability of being the
same person).
Given the rich set of global and local perceptions in human faces, we focus
on perceptual-level mutations in this section.
We first randomly select 100 pairs of face photos from CelebA (image size $64 \times 64$),
then, for each pair, we apply various global/local perceptual mutations (listed in
\T~\ref{tab:eval-face}) to one of the images. Then, we determine the lower and
upper bounds of robustness (i.e., the percentage of inputs that will not change
the prediction). In this setting, if the original prediction is ``yes,''
following GenProver, the computation for the lower/upper bounds is:
\begin{equation*}
  \small
    \begin{aligned}
        \text{Lower Bound} = \sum_i \lambda_i \, \bm{[}\,\forall q \in P_i, q > 0.5\,\bm{]} \\
        \text{Upper Bound} = \sum_i \lambda_i \, \bm{[}\,\exists q \in P_i, q > 0.5\,\bm{]} \\
    \end{aligned}
\end{equation*}
where $\bm{[}\cdot\bm{]}$ outputs $1$ if the input condition holds,
and $0$ otherwise. $P_i$ is the $i$-th segment in outputs of the certified
NN $f$ (recall that the input segment of $f \circ G$ is bent by
non-linear layers in $G$ and $f$), and $\lambda_i$ is the weight of $P_i$ (i.e,
the ratio of its length over the whole length). $q$ is a point in $P_i$.
If the original prediction is ``no,'' users can replace $q > 0.5$ as $q \leq
0.5$.

\begin{table}[!htpb]
    \vspace{-5pt}
    \caption{Quantitative certification for face recognition.}
    \vspace{-10pt}
    \label{tab:eval-face}
    \centering
  \resizebox{0.75\linewidth}{!}{
    \begin{tabular}{l|c|c|c|c}
        \hline
          & Global & \multicolumn{3}{c}{Local} \\
        \cline{2-5}
          & Orientation & Hair & Eye & Nose \\
        \hline
        Upper Bound & 100\% & 98.1\% & 69.7\% & 95.2\% \\
        \hline
        Lower Bound & 97.6\% & 95.0\% & 60.3\% & 90.3\% \\
        \hline
    \end{tabular}
    }
    \begin{tablenotes}
        \footnotesize
        \item 1. Orientation: change face orientation.
        2. Hair: change hair color. 
        \item 3. Eye: open/close eyes, or add
        glasses. 4. Nose: change nose size.
    \end{tablenotes}
   \vspace{-5pt}
\end{table}

\parh{Results.}~As in \T~\ref{tab:eval-face}, the lower/upper bounds vary with
the mutated facial attributes. Notably, orientation is less effective to flip
the prediction, as the NN does not rely on orientation to recognize a face in
most cases. In contrast, mutating eyes is highly effective to deceive the NN.
This is reasonable, as NNs often rely on key facial attributes, such as eyes, to
recognize faces~\cite{li2015convolutional,williford2020explainable}. Hence,
closing eyes or adding glasses will cause NNs to lose crucial information.

\subsubsection{Certifying Autonomous Driving}
\label{subsubsec:eval-drive}

\parh{Setup \& Framework.}~In this evaluation, we incorporate \tool\ into
$\text{AI}^2$~\cite{gehr2018ai2} (i.e., ERAN~\cite{eran}) to certify if the NN
is robust under various severe weather conditions. This is a common setup to
stress NN-based autonomous driving~\cite{zhang2018deeproad,pei2017deepxplore},
and NN's robustness is ensured if it can \textit{retain its steering angle
decisions} under severe weather conditions. We use the ``rainy'' and ``foggy''
filters to simulate the effects of rain and fog. Note that that geometrical and
perceptual-level mutations do not apply here,
because the steering angle is tied to geometrical properties and the road is
frequently altered when mutating perceptions: the ground truth steering decisions
are changed by these mutations.
We randomly select 500 images from the Driving dataset for mutation and
report the percentage of successful certification (i.e., the NN is proved to
retain its prediction) for three degrees of mutations, as in
\T~\ref{tab:eval-driving}. While $\text{AI}^2$ performs \textit{incomplete}
certification, it has SOTA scalability, allowing it to efficiently certify large
images (whose size is $256 \times 128$) in the Driving dataset.

\begin{table}[!htpb]
    \caption{Certification for autonomous driving over severe weather
    conditions. Results are \% of successful certifications.}
    \vspace{-10pt}
    \label{tab:eval-driving}
    \centering
  \resizebox{0.52\linewidth}{!}{
    \begin{tabular}{l|c|c|c}
        \hline
              & Small & Medium & Heavy \\
        \hline
        Rainy & 97.5\% & 80.3\% & 70.9\% \\
        \hline
        Foggy & 96.2\% & 75.2\% & 61.7\% \\
        \hline
    \end{tabular}
    }
\end{table}

\parh{Results.}~\T~\ref{tab:eval-driving} presents the results. The extent
of weather filter is controlled by a floating number weight within $[0, 1]$
(see \S~\ref{subsubsec:style}). To ease analysis, we discretize it into
three degrees (0.2, 0.5, and 0.8) in \T~\ref{tab:eval-driving}. When heavy
weather filters are applied to these images, a considerable number of
certifications fail. This may be a result of the inconsistent predictions (true
positive certification failures) or the over-approximation of layer-propagation
performed in $\text{AI}^2$ (false positive certification failures). Besides, the
``foggy'' filter seems more likely to break consistent predictions of NNs. This
is intuitive, because fog can easily decrease the ``visibility'' of the NN, and
therefore, road details (e.g., the yellow road stripes) for making decisions may
be missing.

\begin{figure}[!ht]
  \vspace{-5pt}
  \captionsetup{font=footnotesize}
  \centering
  \includegraphics[width=0.7\linewidth]{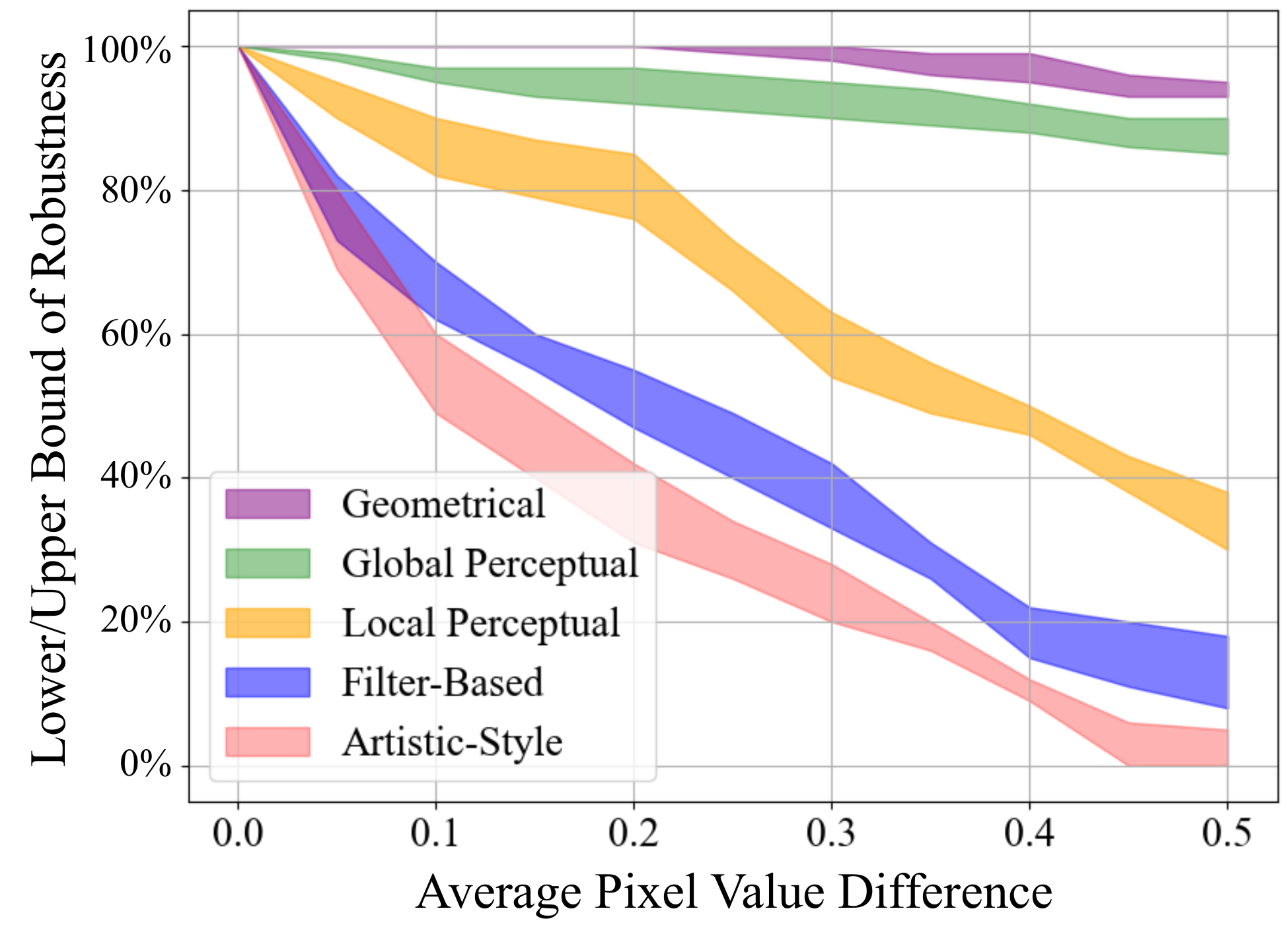}
  \vspace{-10pt}
  \caption{Compare different mutations in terms of lower/upper bounds.}
  \vspace{-10pt}
  \label{fig:eval-comp}
\end{figure}

\subsubsection{Cross-Comparing Different Mutations}
\label{subsubsec:eval-compare}

In this section, we study if the same NN has notably different robustness under
different types of mutations. This step subsumes all mutations
supported by \tool. For a more fine-grained measurement of the robustness, we
incorporate \tool\ into GenProver for quantitative certification.

\parh{Dataset \& NN.}~We use CelebA, as the resolution of CIFAR10/MNIST is
insufficient for fine-grained local perceptual mutations. We still
certify face recognition but for a ResNet~\cite{he2016deep}, which is one highly
successful and pervasively-used NN for computer vision tasks in recent 
years.

\parh{Setup.}~To compare the extent of these mutations in a unified way, we use
the Average Pixel value Difference (APD):
\begin{equation*}
  \small
  APD = \frac{\sum_i | x_i - x_i' | \,\bm{[}\,x_i \neq x_i'\,\bm{]}}
  {\sum_i \bm{[}\,x_i \neq x_i'\,\bm{]}}
\end{equation*}
where $x$ is the original image and $x_i$ is its $i$-th pixel. $x'$ is
the mutated image. $\bm{[}\cdot\bm{]}$ yields $1$ if $x_i \neq x_i'$ and $0$
otherwise. Note that APD is only calculated over changed pixels to ``normalize''
the value, because geometrical and local-perceptual mutations do not change all
pixels.

\parh{Results.}~\F~\ref{fig:eval-comp} depicts how the lower/upper bounds of
robustness change with varying APD values across different mutations. It is
seen that for well-trained NNs like ResNet, geometrical mutation is
\textit{not} a major concern to impede its robustness: extensive geometrical
mutations (when APD scores are high) can hardly flip predictions, as implied in
\F~\ref{fig:eval-comp}. 
In contrast, stylized mutations, including both artistic-style and filter-based
mutations, are effective at challenging NNs. This result is consistent to the
recently-observed ``texture-bias'' in common NNs
networks~\cite{geirhos2018imagenet,hermann2020origins}. In particular, people
have observed that NNs heavily rely on texture to recognize objects, and that
the predictions can be easily changed by mutating the texture (mostly by
artistic-style transfer).
Local-perceptual mutation has a major impact on the robustness as well. In
contrast, global-perceptual mutation is less effective to stress the NNs.
Recall that, as discussed in \S~\ref{subsubsec:eval-face}, NNs
typically rely on key attributes (such as eyes) for predictions. While
local-perceptual mutations like ``closing eyes'' stress the target NNs
by changing those attributes, they are mostly retained in global-perceptual and
geometrical mutations.

\section{Discussion and Extension}
\label{sec:discussion}

In general, extending \tool\ can be conducted from the following aspects.

\parh{Mutations:}~\tool\ is data driven, in the sense that to add new
semantic-level mutations, users only need to add images with new semantics to
$G$'s training data. As evaluated in \Appx~\ref{appx:ablation}, due to the
continuity, \tool\ can infer ``intermediate images'' between the original and
the mutated images without having them in the training data.
According to our empirical observations, only dozens of data are enough
to add one new mutation.

\parh{Data:}~Generative models have been widely adopted for generating data of
various forms, such as text~\cite{zhao2018adversarially},
audios~\cite{donahue2018adversarial}. \tool's technical pipeline is independent
of the implementation of generative models; therefore, \tool\ can support other
data forms by employing the corresponding $G$.

\parh{Objectives/Tasks:}~\tool\ can boost certification of other NN tasks
(e.g., object detection) and objectives (e.g., fairness), as those tasks also
accept images as inputs and fit semantic-level mutations studied in this paper.
For example, fairness certifications typically assess whether certain semantics
(which are typically perceptual semantics such as gender) affect the NN
prediction~\cite{urban2020perfectly,ruoss2020learning}. \tool\ can boost them
since it enables independent and analysis-friendly perceptual mutations.
In short, despite the fact that supporting other tasks/objectives is not \tool's
main focus, such extensions should be straightforward, since \tool\ is not
limited to particular certification techniques/objectives.

\section{Conclusion}
\label{sec:conclusion}

We have proposed \tool\ which enables the certification of NN robustness under
diverse semantic-level image mutations. We identify two key properties,
independence and continuity, that result in a precise and analysis-friendly
input space representation. We show that \tool\ empowers de facto complete,
incomplete, and quantitative certification frameworks using semantic-level
image mutations with moderate cost, and under various real-world scenarios.

\section*{Acknowledgement}

We thank all anonymous reviewers and our shepherd for their
valuable feedback. We also thank authors of~\cite{li2023sok,balunovic2019certifying,
mirman2021robustness,sotoudeh2019computing,gehr2018ai2,eran,li2021tss}
for their high-quality and open-sourced tools, which greatly help us
set up experiments in this paper.

The HKUST authors were supported in part by the HKUST 30 for 30 research
initiative scheme under the contract Z1283.

\bibliographystyle{plain}
\bibliography{bib/main}

\begin{appendix}

\section{Proof for Continuity (\E~\ref{equ:continuous})}
\label{appx:proof}

This section proves that continuity (\E~\ref{equ:continuous}) is satisfied,
by regulating the generative model with \E~\ref{equ:objective} as in
\A~\ref{alg:continuity}.
\E~\ref{equ:objective} limits the gradients of $G$ over $z \in \mathcal{Z}$. As
a result, $\| \mathbf{J}(z) \|$ is bounded. Suppose its maximum and minimum
values are $A$ and $\frac{1}{B}$, respectively. For any path on $\mathcal{Z}$,
namely, $\gamma_{\mathcal{Z}}(t): [0, t] \rightarrow \mathcal{Z}$, the length of
the corresponding curve on $\mathcal{M}$ is

\begin{equation}
  \mathrm{Len}(G(\gamma_{\mathcal{Z}}(T)))
  = \int_0^T \norm{
    \mathbf{J}(z_t) \dot{\gamma}_{\mathcal{Z}}(t)
  } \mathrm{d}t
\end{equation}

\noindent Let $z$ and $z'$ be the points at time $0$ and $T$, respectively.
Since by design $\mathcal{Z}$ follows a continuous distribution, we have
$\mathrm{Len}(\gamma_{\mathcal{Z}}(T)) = l_{\mathcal{Z}} (z, z')$. And
therefore, the following inequation holds:

\begin{equation}
  \mathrm{Len}(G(\gamma_{\mathcal{Z}}(t)))
  \leq A \int_0^T \norm{\dot{\gamma}_{\mathcal{Z}}(t)} \mathrm{d}t
  = A \cdot l_{\mathcal{Z}} (z, z').
\end{equation}

\noindent where $l_{\mathcal{Z}} (z, z')$, the distance between $z$ and $z'$,
equals to the length of the shortest path connecting $z$ and $z'$. Thus, we have

\begin{equation}
  \frac{1}{B} \cdot l_{\mathcal{Z}} (z, z')
  = \frac{1}{B} \int_0^t \norm{\dot{\gamma}_{\mathcal{Z}}(t)} \mathrm{d}t
  \leq l_{\mathcal{M}} (G(z), G(z'))
\end{equation}

\noindent Similarly, since $l_{\mathcal{M}} (G(z), G(z'))$ equals to the
shortest length of the paths between $G(z)$ and $G(z')$ on $\mathcal{M}$, we
have $l_{\mathcal{M}} (G(z), G(z')) \leq
\mathrm{Len}(G(\gamma_{\mathcal{Z}}(t)))$.

Let $C = \max(A, B)$, the following equation thus holds:

\begin{equation}
  \frac{1}{C} \cdot l_{\mathcal{Z}} (z, z')
  \leq l_{\mathcal{M}} (G(z), G(z'))
  \leq C \cdot l_{\mathcal{Z}} (z, z').
\end{equation}

Besides the theoretical analysis, we also empirically calculate the value of
$C$ (which $\geq 1$) over 10K inputs for generative models trained on different
datasets. Results are in \T~\ref{tab:c-value}. We interpret the results
as generally promising and intuitive.

\begin{table}[!htpb]
  \caption{Values of $C$ in \E~\ref{equ:continuous}.}
  \vspace{-5pt}
  \label{tab:c-value}
  \centering
\resizebox{0.6\linewidth}{!}{
  \begin{tabular}{c|c|c|c}
      \hline
       MNIST & CIFAR10 & CelebA & Driving \\
      \hline
      2 & 5 & 3 & 3 \\
      \hline
  \end{tabular}
  }
\end{table}
\vspace{-10pt}

\section{Cost Analysis}
\label{appx:cost}

In practice, $f \circ G$ will typically double the number
of layers (\#layers) in $f$, as \#layers is correlated with input image
resolution: if $f$ is deeper (meaning its inputs exhibit higher resolution),
then $G$ should also be deeper to generative more fine-grained images; the same
holds true if $f$ is much shallower. 
As explained in \S~\ref{sec:overview}, if $f$ has $L$ layers and $N$ neuron width
(i.e., the maximum number of neurons in a layer),
the conventional cost for complete certification is $O((2^{N})^{L})$~\cite{li2023sok}.
In \tool, the \#layers of $f \circ G$ is approximately $2L$. Therefore, based on
the results of~\cite{sotoudeh2019computing}, \tool's cost is
$O({N}^{2L}) = O({({N}^2)}^L)$ --- \tool\ reduces an exponential cost
$O(2^N)$ to a polynomial $O({N}^2)$ for a given $L$. As evaluated in \T~\ref{tab:eval-bound} and
\T~\ref{tab:eval-class}, \tool's cost is empirically moderate.

Similarly, the cost for conventional incomplete certification is $O(NL)$~\cite{li2023sok}.
Since $f \circ G$ doubles the number of layers, \tool\ increases the certification
cost to $O(2NL)$, which is still polynomial to neuron width and the
number of layers.

\section{Ablation Studies}
\label{appx:ablation}

As mentioned in \S~\ref{sec:motivation}, $G$ enables ``data-driven'' mutations.
Therefore, at this step we launch controlled experiments to explore how the
training data of $G$ would affect the enabled mutations. More specifically, we
study how the mutation quantity and extent are affected.

We observe that for CIFAR10 (image size is $32 \times 32$), only approximately
2$\sim$4 global perceptual mutations can be identified. In comparison to
the 35 global perceptual mutations in CelebA-HQ (image size is $256 \times 256$), we
interpret that the resolution of training images has a notable impact on the
diversity of perceptual mutations\footnote{
    For local perceptions, the selected local region also affects
    the number of mutations. In general, face photos in CelebA(-HQ) should
    support more local perceptual-level mutations given the rich
    perceptions in human faces.
}. This is reasonable since, with a higher
resolution, more details of perceptual contents can be modeled by the generative
models.

In practice, it is infeasible to automatically decide if a specific perceptual
mutation exists or to what extent it is applied. Thus, we focus on geometrical
mutations in the following controlled experiments. Nevertheless, geometrical
properties can be affected by diverse semantics in real datasets (e.g., a digit
zero is written in different scales). To avoid such side effects, we construct a
synthetic dataset of rectangle images (scripts are in~\cite{snapshot}). In
short, we first prepare a seed image which has one square with one-length sides
(``a unit square'') at the center. We then apply different geometrical
mutations over the seed  to generate training data for $G$. This way, we can
precisely control the number/extent of geometrical mutations involved in the
training data.

We summarize the results as follows. First, the mutation supported by $G$ is
\textit{not} arbitrary: $G$ can only enable mutations whose mutated properties
exhibit observed variations in the training data. Second, the maximal variation
of the target property determines the largest extent (i.e., $\| \delta_{\max}
\|$) of one mutation. More importantly, we emphasize that it is
\textit{unnecessary} to cover all possible $\| \delta \|$ values in the training
data in order to enable all mutations within $\| \delta \| \leq \| \delta_{\max}
\|$. Since $\mathcal{Z}$ follows a continuous distribution (introduced in
\S~\ref{sec:motivation}) and $G$ satisfies the continuity (\S~\ref{subsec:continuity}),
only a few $\| \delta \|$ values within $\| \delta \| \leq \| \delta_{\max} \|$
need to be covered in the training data. This roots from the manifold
hypothesis~\cite{bengio2013representation,zhu2018image} and demonstrates the
merit of our generative model-based approach, as introduced in
\S~\ref{sec:motivation}.
\end{appendix}

\end{document}